\documentclass[useAMS,usenatbib,dvips]{mn2e}
\usepackage{graphicx}

\newcommand{\magcir}{\raise
-2.truept\hbox{\rlap{\hbox{$\sim$}}\raise5.truept
\hbox{$>$}\ }}
\newcommand{\minmag}{\raise-2.truept\hbox{\rlap{\hbox{$<$}}\raise
6.truept\hbox
{$>$}\ }}

\newcommand{\lya}{Ly$\alpha$~}
\newcommand{\lyb}{Ly$\beta$~}

\newcommand{\be}{\begin{equation}}
\newcommand{\ee}{\end{equation}}
\newcommand{\ba}{\begin{eqnarray}}
\newcommand{\ea}{\end{eqnarray}}
\newcommand{\brr}{\begin{array}}
 
\newcommand{\err}{\end{array}}
\newcommand{\bc}{\begin{center}}
\newcommand{\ec}{\end{center}}

\newcommand{\hm}{\,h^{-1}{\rm Mpc}}

\newcommand{\kms}{km s$^{-1}$}
\newcommand{\eff}{$\tau_{\ion{H}{i}}^{\rm eff}$}

\DeclareMathAlphabet{\mathsc}{OT1}{cmr}{m}{sc}
\def\testbx{bx}%
\DeclareRobustCommand{\ion}[2]{%
\relax\ifmmode
\ifx\testbx\f@series
{\mathbf{#1\,\mathsc{#2}}}\else
{\mathrm{#1\,\mathsc{#2}}}\fi
\else\textup{#1\,{\mdseries\textsc{#2}}}%
\fi}

\title[The flux distribution of the \lya forest]
{An improved measurement of the flux distribution of the \lya forest in QSO
absorption spectra: the effect of  continuum fitting, metal contamination and  noise 
properties\thanks{Based on data taken from ESO archive
obtained with UVES at VLT, Paranal, Chile.}}
\author[Kim et al.] 
{T.-S. Kim$^{1, 2}$, 
J. S. Bolton$^{2, 3}$, M. Viel$^{2,4}$, 
M. G. Haehnelt$^{2}$, R. F. Carswell$^{2}$
\\ 
$^1$ Astrophysikalisches Institut Potsdam, 
An der Sternwarte 16, D-14482 Potsdam, Germany\\
$^2$ Institute of Astronomy, Madingley Road, Cambridge 
CB3 0HA, UK\\
$^3$ Max-Planck-Institut f\"ur Astrophysik, Karl-Schwarzschild-Str. 1,
85748 Garching bei M\"unchen, Germany\\
$^4$ INAF-Osservatorio Astronomico di Trieste, Via G. B. Tiepolo 11,
I-34131 Trieste, Italy\\
\\}

\begin{document}

\date{Received 2007 July 7; Accepted 2007 September 3}

\maketitle

\begin{abstract}
We have performed an extensive Voigt profile analysis of the neutral
hydrogen (\ion{H}{i}) and metal absorption present in a sample of 18 high
resolution, high signal-to-noise QSO spectra observed with VLT/UVES.
We use this analysis to separate the metal contribution from the
\ion{H}{i} absorption and present an improved measurement of the flux
probability distribution function (PDF) due to  \ion{H}{i} absorption
alone at $<\!z\!>\,=$ 2.07, 2.52, and 2.94.   The flux PDF is
sensitive to the continuum  fit in the normalised flux range $0.8 < F
< 1.0$ and to metal absorption at $0.2 < F < 0.8$.   Our new
measurements of the flux PDF due to \ion{H}{i} absorption alone are
systematically lower at $0.2 < F < 0.8$ by up to 30\% compared to the
widely used measurement of \cite{mc00}, based on a significantly
smaller sample of Keck/HIRES data. This discrepancy is  probably 
due to a combination of our improved removal of the metal absorption
and cosmic variance, since variations in the flux PDF between
different lines-of-sight are large.
The \ion{H}{i} effective optical depth $\tau_{\ion{H}{i}}^{\rm eff}$  
at $1.7 < z < 4$ is best fit with a
single power law, $\tau_{\ion{H}{i}}^{\mathrm{eff}} =  
(0.0023 \pm 0.0007) (1+z)^{3.65 \pm 0.21}$, 
in good agreement with  previous
measurements from comparable data.  As also found previously, the
effect of noise on  the flux distribution is not significant in high
resolution,  high signal-to-noise data.
\end{abstract}


\section{Introduction}

The Ly$\alpha$ forest refers to the numerous, narrow absorption
features bluewards of the Ly$\alpha$ emission line in the spectra of
high-redshift QSOs.  Our understanding of the origin of the Ly$\alpha$
forest has made major progress with the advent of high resolution
spectroscopy \citep{cowie95,hu95,lu96,kirk97,rau97,cri00,kim01,jan06}
and the availability of hydrodynamic cosmological simulations
\citep{cen94,miralda96,zhang97,weinberg98,
cro98,theuns98,bryan99,dave99,cro02,jena05}.  Comparison of the
predictions from hydrodynamic  cosmological simulations with
observations have established a picture where the Ly$\alpha$ forest is
due to the neutral hydrogen (\ion{H}{i}) component of the warm ($\sim
10^4$K) photoionised intergalactic medium (IGM) which traces
moderate-amplitude density  fluctuations of the  (dark) matter
distribution in a simple manner. In this picture the spectra of
high-$z$ QSOs provide a one-dimensional probe of the matter  density
along the line-of-sight to high redshift QSOs.  The flux distribution
in QSO absorption spectra   thus encodes  information on the
underlying matter  distribution and evolution and the  Ly$\alpha$
forest has been recognised as a powerful cosmological tool,
complementing other cosmological  probes.

So far it is mainly the power spectrum of the flux distribution which
has been used to quantitatively probe the properties of the large
scale distribution of matter and its dependence on cosmological
parameters. The flux power spectrum is sensitive to the matter power
spectrum over a wide redshift range ($2<z<4$) and on scales of
1--$50\hm$, significantly smaller than those probed by CMB fluctuations,
galaxy surveys and gravitational lensing (Croft et al. 2002; Seljak,
McDonald \& Makarov 2003; Viel, Haehnelt \& Springer 2004b; McDonald
et al. 2006; Viel, Haehnelt \& Lewis 2006; Lesgourges et al. 2007).

The probability distribution of the flux is also sensitive to the
spatial distribution of dark matter and cosmological parameters, in
particular, the amplitude of matter fluctuations
\citep{rau97,weinberg98,theuns00,mek01}.  However, the effect of the
thermal state of the IGM (Theuns et al. 2000) and  other uncertainties
make it more difficult to extract this information from the flux
PDF. First attempts at a joint analysis of the flux PDF and the flux
power spectrum have been made \citep{des05, lid06, des07}. The
analysis of Desjaques et al.  gives a somewhat smaller amplitude for
the matter power spectrum compared to studies based on the flux power
spectrum alone. Note, however, that these results were based on
dark-matter-only numerical simulations in which the gas distribution
was modelled in an approximate way rather than on full hydrodynamical
simulations.

The measured flux PDF is sensitive not only to the \ion{H}{i}
absorption, but also, unfortunately, to continuum level uncertainties,
contaminating heavy element absorption, and noise.  The effect of
continuum fitting uncertainties on the \ion{H}{i} effective optical
depth has been discussed in detail \citep{pre93, kim02, ber03, tyt04,
kirk05}, but their impact on the flux PDF has not been considered as
extensively (see Rauch et al. 1997 and McDonald et al. 2000 for
notable exceptions).  Very little, if anything, has been done to
quantify the effect of metal absorption and noise on the flux PDF.  To
remedy these omissions, we perform a detailed Voigt profile analysis
of the metal and \ion{H}{i} absorption in 18 high resolution, high
signal-to-noise (S/N) VLT/UVES QSO absorption spectra to assess the
impact of continuum fitting uncertainties, metal absorption, and noise
properties on the \ion{H}{i} flux probability distribution 
at $1.7 < z < 3.2$ and the \ion{H}{i} effective
optical depth at $1.7<z<4$.
 
Our study is based on a factor of two larger sample and probes towards
lower $z$ than similar work based on 8 Keck/HIRES spectra by
\citet{mc00}. It also differs significantly in the way we perform the
continuum fitting and, more importantly, in the level of the
characterisation of metal absorption.  Most previous estimates of the
effective optical depth and the flux PDF due to \ion{H}{i} absorption
have either removed the metal lines {\it statistically} \citep{tyt04,
kirk05}, excluded the metal-contaminated regions from the study
\citep{mc00, lid06} or included absorption due to metal lines
\citep{son04}.

We have carefully identified and {\it fitted} metal and \ion{H}{i}
absorption lines with Voigt profiles for our sample of 18 spectra. The
line fitting is essentially complete in the Ly$\alpha$ forest region
and redwards of the Ly$\alpha$ emission lines, but less complete in
the higher-order Lyman forest where the corresponding spectral regions
were not always available.  During the fitting process, we have
adjusted the initially estimated continuum level by a localised
continuum fit in order to get an overall satisfactory fit to
multiple-transition metal lines and higher-order Lyman series lines.
This approach enables us to remove the metal contribution in the
forest {\it directly} and to recover one long continuous metal-free
forest spectrum, and also to obtain an improved estimate of the continuum
level.

The paper is organised as follows.  In Section \ref{sec2}, we describe
the data used for our analysis.  In Section \ref{sec3}, we present the
Voigt profile fitting analysis used to remove the metal contribution
in the forest.  We investigate systematics of uncertainties affecting
the flux PDF in Section \ref{sec4}.  The observed flux PDF function
and effective optical depth are presented in Section \ref{sec5}.  Our
conclusions are given in Section \ref{sec6}.

\section{The Data}
\label{sec2}

\subsection{Description of the sample} 

\begin{figure}
\includegraphics[width=9cm]{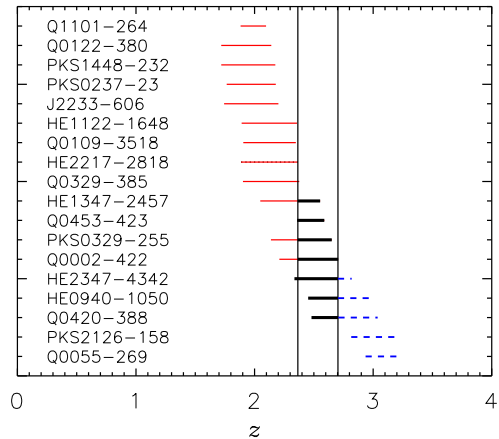}
\vspace{-1.8cm}
\caption{The redshift range of the \lya\ forest in the spectra used for
this study. The sample is divided into redshift bins.  The thin dotted
lines, thick solid lines and dashed lines represent the
$<\!z\!>\,=2.07$, $<\!z\!>\,=2.52$ and $<\!z\!>\,=2.94$ sub-sample,
respectively.}
\label{fig1}
\end{figure}

We have selected a sample of 18 spectra from the LUQAS sample 
\citep{kim04} of 27 spectra obtained with
VLT/UVES \citep{dekker00}. 
The spectra in LUQAS (Large sample of
UVES QSO Absorption Spectra) were taken from the ESO archive and a
large fraction of them were observed as part of the Large Program
166.A-0106 \citep{ber}.  The observations were performed in the period
1999--2004.  All spectra have high signal-to-noise (S/N $\ge$ 25--30)
in the \lya forest region and a resolution of $R \sim 45\,000$. The
wavelengths are helio-centric and vacuum-corrected. The spectra were
sampled with pixels of width 0.05 \AA\/.

From the 27 spectra in the LUQAS sample, we have chosen those with a
S/N in the forest region of at least 30--50, a full coverage of the
Ly$\alpha$ forest and at least some coverage of the Ly$\beta$
forest. Most of the 18 spectra cover the wavelength range of the
higher order Lyman forest down to the atmospheric cut-off at
3050 \AA\/.  Some, however, cover only a part of the \lyb forest due to
their low emission redshift. For \lya absorption at $z \le 1.99$ the
corresponding Ly$\beta$ absorption is not covered in these optical
spectra. Note that we did not include the spectrum of HE1341$-$1020
which would have met these criteria but is a mini-BAL.

In Table~\ref{tab1} we list the characteristic properties of the 18
QSO spectra in our sample.  The signal-to-noise ratio is per pixel
with the first (second) number corresponding to the lower (upper)
limit of the wavelength range analysed.  Note that the S/N is not
constant across the spectrum. To avoid the proximity effect, we
excluded a region corresponding to 4,000 \kms\/ bluewards of the \lya
emission line.  Lyman limit absorption systems (LLSs) with column
densities $10^{17.2} \mathrm{cm}^{-2} \le N_{\ion{H}{i}} < 10^{19}
\mathrm{cm}^{-2}$ are included in our flux distribution study, while
50 \AA\/ segments of the spectra have been excised to the left and
right of the centre of sub-damped Ly$\alpha$ absorption systems
(sub-DLAs, $N_{\ion{H}{i}} = 10^{19-21.3} \mathrm{cm}^{-2}$). None of
our spectra contain a damped Ly$\alpha$ system ($N_{\ion{H}{i}} \ge
10^{21.3} \mathrm{cm}^{-2}$) in the observed wavelength range down to
either the atmospheric cutoff or at the Lyman limit discontinuity
caused by LLSs or sub-DLAs along the spectra with non-zero observed
fluxes.

The Ly$\alpha$ forest regions of the spectra were grouped into 3
redshift bins with median redshifts $<\!z\!>\,=$ 2.07, 2.52 and 2.94,
as shown in Fig.~\ref{fig1} and listed in Table~\ref{tab_a} in
Appendix A. We have chosen these redshift bins as a compromise between
minimising the Poisson fluctuations of the statistical quantities we
are interested in and being able to investigate their redshift
evolution.  As we will show in section \ref{sec5:1}, spatial
fluctuations in the flux PDF are large and substantial redshift paths
are necessary to reduce the the Poisson {\it noise} from individual
absorption systems.  If an object contains an absorption profile which
straddles a redshift  bin boundary we have assigned it to the redshift
bin in which most of it lies, and adjusted the wavelength range
contributed by that QSO accordingly.

The spectra used here are essentially the same as those used 
in \citet{kim04} in the analysis of the LUQAS sample. There are, 
however, three differences  between the spectra in the 
LUQAS sample and the spectra analysed here: 

First, additional data for HE1122$-$1648, HE2347$-$4342 and
PKS2126$-$158 has been obtained since the LUQAS data reduction.  The
additional data has been reduced with MIDAS/UVES as described in
\citet{kim04}, and added using the UVES\_popler (UVES POst PipeLine
Echelle Reduction)
program{\footnote{http://astronomy.swin.edu.au/mmurphy/UVES\_popler.html,
kindly provided by Dr. M. Murphy.}, resulting in significantly larger
S/N for the spectra of these three QSOs.

Second, we have re-reduced and performed a first-order flux
calibration for the spectra of four QSOs (Q0420$-$388, HE0940$-$1050,
Q0002$-$422 and HE2217$-$2818) in order to estimate the normalisation
uncertainty between calibrated spectra and un-calibrated spectra.  The
UVES spectra do not normally have an absolute flux calibration, due to
the lack of well-calibrated standard stars with spectra taken at the
resolution of the instrument{\footnote{The UVES pipeline reduction
program determines the large-scale continuum shape from the ratio of
the spectral shape of the target and the flat-field flux. For a fixed
instrumental setting, the extracted spectra of the same object show a
more or less similar shape of the large-scale continuum.}}.
Consequently, most spectra in the LUQAS sample were not
flux-calibrated. Even in the few cases where flux calibrations were
performed, such as for HE2217$-$2818, these were done echelle order by
echelle order and are far from perfect. It is not uncommon for a
standard star to have strong hydrogen Balmer series with a width
larger than a echelle order ($\sim 25$ \AA\/), and then the spectrum
cannot be properly flux-calibrated in this way.  A lack of a proper
flux calibration and deficiencies in the flat-fielding procedure
sometimes leave unwanted, oscillatory features in the extracted
spectrum, with periods corresponding to the scale of one echelle
order.  The flux is then higher at the central wavelength and lower at
both ends of a given echelle order compared to the flux at the same
wavelengths from adjacent echelle orders. In addition, the
non-calibrated spectra show a different continuum shape for the same
wavelength regions covered by the different CCDs.  These deficiencies
aggravate the difficulties of continuum fitting the spectra,
especially at high redshifts.  Instead of an order-by-order flux
calibration, we performed a first-order flux calibration as follows.
We corrected each merged science spectrum using the master response
function{\footnote{http://www.eso.org/observing/dfo/quality/UVES/qc/std\_qc1.html}}
provided by ESO for a given observing period.  When the master
response functions were not available for a particular period, we
scaled the un-calibrated spectra to the flux-calibrated spectra from
other periods. The flux-calibrated spectra were then combined as
described in \citet{kim04}.

Third, and most importantly, we performed new continuum fits to all
spectra in the sample optimised for a Voigt profile analysis.   The
normalised LUQAS spectra and the normalised spectra analysed here
therefore differ slightly.  Details of the continuum fitting are
described in the next section.

\begin{table*}
\caption{Analysed QSOs}
\label{tab1}
\begin{tabular}{llccccrl}
\hline
\noalign{\smallskip}
QSO & $z_{\mathrm{em}}^{\rm a}$ & $z_{\mathrm{Ly\alpha}}$ & $\lambda_{\mathrm{Ly\alpha}}(\rm{\AA})$
& S/N & Continuum (\%)$^{\rm b}$  & Lyman limit$^{\rm c}$ & notes$^{\rm d}$ \\
\noalign{\smallskip}
\hline
\noalign{\smallskip}
Q0055--269          & 3.655$^{\rm e}$ & 2.936--3.205 & 4785--5112 & 80--50 & 1$-$3 & 2288 &   \\ 
PKS2126--158        & 3.279 & 2.815--3.205 & 4638--5112 & 50--200 & 1 &3457$^{}$ & two sub-DLAs at $z=$2.768 \& 2.638 \\ 
Q0420--388          & 3.116$^{\rm e}$ & 2.480--3.038 & 4231--4909 & 100--140  & 1 & 3754  & a sub-DLA at $z=$3.087\\
HE0940--1050        & 3.078 & 2.452--3.006 & 4197--4870 & 50--130 & 1$-$2 & $\le 3200$ & \\
HE2347--4342$^{\rm f}$  & 2.874$^{\rm e}$ & 2.336--2.819 & 4055--4643 & 100--160 & 1$-$2 & ... & multiple associated systems \\
Q0002--422          & 2.767 & 2.209--2.705 & 3901--4504 & 60--70 & 2 & 3025 &  \\
PKS0329--255        & 2.704$^{\rm e}$ & 2.138--2.651 & 3815--4439 & 30--55 & 2$-$4 & 3157 & associated system at 4513.7 \AA\\
Q0453--423          & 2.658$^{\rm e}$ & 2.359--2.588 & 4084--4362 & 90--100 & 1$-$2 & 3022 & a sub-DLA at $z=$2.305 \\
                    &             & 2.091--2.217 & 3758--3911 & 60--100 & 1$-$2  &  & only for the \ion{H}{i} opacity    \\ 
HE1347--2457        & 2.609$^{\rm e}$ & 2.048--2.553 & 3705--4319 & 85--100 & 2 & 2237 & \\
Q0329--385          & 2.434 & 1.902--2.377 & 3528--4105 & 50--55 & 2$-$3 &  $\le 3050$ & \\ 
HE2217--2818        & 2.413 & 1.886--2.365 & 3509--4091 & 65--120 & 1$-$2 & 2471 & \\ 
Q0109--3518         & 2.405 & 1.905--2.348 & 3532--4070 & 60--80  & 2 & 2163 & \\
HE1122--1648        & 2.404 & 1.891--2.358 & 3514--4082 & 70--170 & 1$-$2 & $\le 1629$ & \\
J2233--606          & 2.250 & 1.756--2.197 & 3335--3886 & 30--50  & 2$-$4 & $\le 1750$ & \\
PKS0237--23         & 2.223$^{\rm e}$ & 1.765--2.179 & 3361--3865 & 75--110 & 1$-$2 & $\le 3050$ 
                    & a sub-DLA at $z=$1.673 \\ 
PKS1448--232        & 2.219 & 1.719--2.175 & 3306--3860 & 30--90 & 2$-$4 & $\le 3050$ & \\
Q0122--380          & 2.193 & 1.700--2.141 & 3282--3819 & 30--80 & 2$-$4 & $\le 3052$ & \\
Q1101--264          & 2.141 & 1.880--2.097 & 3503--3765 & 80--110 & 1$-$2 & 2597 & a sub-DLA at $z=$1.839 \\
                    &       & 1.659--1.795 & 3233--3398 & 45--75  & 2$-$3 &  &  only for the \ion{H}{i} opacity   \\                 
\noalign{\smallskip}
\hline
\end{tabular}
\begin{list}{}{}
\item[$^{\mathrm{a}}$]
The redshift is measured from the observed \lya emission line of the QSOs. 
The redshift based on the emission lines is known to be under-estimated 
compared to the one measured from the absorption lines of the host galaxies
\citep{tyt92, berk01}.
\item[$^{\mathrm{b}}$]
Rough estimate of the  continuum fitting uncertainty. It varies across the spectrum. 
\item[$^{\mathrm{c}}$]
The wavelength of the Lyman limit for each spectrum is defined here as
the wavelength below which the observed flux is zero. This does not necessarily 
correspond to the  Lyman limit of an identified  Lyman limit system/sub-damped
Ly$\alpha$ system. Whenever a FOS/HST or STIS/HST spectrum is not available,
the Lyman limit is assumed to occur at the shortest observed optical
wavelength.  FOS/STIS spectra from the HST archive are available for 
the following QSOs (any snap-shot survey or spectra with very low S/N are not used):
Q0055$-$269: FOS (PI: Burbidge, Proposal ID: 3199);        
HE2347$-$4342: STIS (PI: Heap, Proposal ID: 7575; PI: Kriss, Proposal ID:
8875), FOS (PI: Reimers, Proposal ID: 6449), GHRS (PI: Reimers, Proposal ID: 6449);
Q0002$-$422: FOS (PI: Rao, Proposal ID: 6577);
Q0453$-$423: FOS (PI: Rao, Proposal ID: 6577);
HE1347$-$2457: STIS (PI: Webb, Proposal ID: 9187);
HE2217$-$2818: STIS (PI: Webb, Proposal ID: 9187); 
Q0109$-$3518: STIS (PI: Webb, Proposal ID: 9187);
HE1122$-$1648: STIS (PI: Baldwin, Proposal ID: 9885), FOS (PI: Reimers, Proposal
ID: 5950); 
J2233$-$606: STIS (PI: Williams, Proposal ID: 8058);
Q1101$-$264: FOS (PI: Bahcall, Proposal ID: 5664)
\item[$^{\mathrm{d}}$]
In spectra with a sub-damped Ly$\alpha$ system, we discard at least
50 \AA\/ on each side of the centre of the sub-DLA system.  
\item[$^{\mathrm{e}}$] 
The emission redshift is uncertain due  to  absorption systems 
at the peak of the Ly$\alpha$ emission line or the occurrence of
multiple peaks.
\item[$^{\mathrm{f}}$]
The spectrum shows very strong \ion{O}{vi} absorption blended with  two
saturated Ly$\alpha$ absorption systems at 4012--4052 \AA\/ \citep{fech04}. 
Since the line parameters
for these Ly$\alpha$ systems cannot be well constrained (their 
corresponding Ly$\beta$ is below the partial Lyman limit produced by the $z \sim 2.738$ systems),
we discarded this wavelength region.
\end{list}
\end{table*}

\subsection{Continuum fitting} 
\label{2norm}

The absorption due to intervening neutral hydrogen and metals  is
imprinted on the QSO emission spectrum, which has several, normally
broad,  emission lines overlaid on continuum emission from thermal and
non-thermal radiation processes. The spectral energy distribution
(SED) of the emission therefore varies significantly between different
QSOs.

Most statistical analysis of the absorption thus require the fitting
of a {\it continuum} characterising the SED of the unabsorbed emission
of the QSO (see \cite{lid06} for a recent study that uses spectra
that are not normalised in this way). Careful continuum fitting is
particularly important for a Voigt profile analysis of QSO absorption
spectra such as the one performed here. Estimating the unabsorbed emission
is particularly difficult at high redshifts where an increasing
fraction of the spectrum shows significant absorption.  The problem
of continuum fitting has been extensively discussed in the literature
\citep{kim02, ber03, sel03, son04, tyt04, kirk05, lid06} and we only
summarise a few important points here.

First it should be noted that there are fundamental differences in the
way continuum fitting is performed for low/intermediate resolution,
low S/N data such as the SDSS \lya forest data and high resolution,
high S/N data obtained with high resolution spectrographs such as
VLT/UVES and Keck/HIRES.  For low/intermediate resolution, low S/N
data the unabsorbed continuum level is normally estimated using a
simple extrapolation from the much less absorbed region of the
spectrum redward of the \lya emission line at wavelengths
$\lambda_{\mathrm{rest-frame}} \ge 1250$ \AA\/. The continuum is
thereby assumed to be a power law with some superimposed emission
lines (Press et al. 1993; Bernardi et al. 2003).

For high resolution, high S/N spectra the continuum is normally fitted
by {\it locally} connecting apparently absorption-free regions
\citep{mc00, kim04, kirk05}. This detailed local fitting of a
continuum is time consuming and ceases to work well once the redshift
of the QSO becomes large ($z_{\mathrm{em}} > 4$), where severe
blending makes it impossible to identify any un-absorbed regions. In
the redshift range considered here, $1.5 < z_{\mathrm{em}} < 3.5$,
however, this method still works well.  We used it to determine an
initial guess for the continuum level for the 18 spectra in our
sample.  Due to the high resolution, high S/N of our data, emission
lines both weak and strong (mainly near 1073 \AA\/ and 1123 \AA\/ in
rest-frame; Bernardi et al. 2003; Tytler et al. 2004) are easily
identified in the forest, despite the lack of an appropriate flux
calibration. The strong, broad ozone absorption bands at $\le $ 3400 \AA\/
only affect the \lya forest at the lower end of the redshift
range of our sample \citep{sch91}.  However, this is compensated for by
the rather low density of absorption features at low redshift which
leaves more absorption-free regions to fit the continuum in our high
resolution, high S/N spectra.

Most of the spectra were not flux-calibrated, and the flat-fielding
procedure was not always ideal, and as a consequence there were some
spurious broad features in a number of the spectra. Rather than attempt
to fit the entire spectrum using an impractically high order function,
the spectrum was divided into chunks. The size of these was determined
interactively, and depended on the redshift and the presence of strong
absorption systems. Typically these chunks were 150--300 \AA\/ long in
the Ly$\alpha$ forest part of each spectrum, and at longer wavelengths
ranged from $\sim 200$ \AA\/ in QSO emission lines up to $\sim 1000$ \AA\/
elsewhere. Legendre polynomials of order $\sim 20$ in the Ly$\alpha$ forest,
and up to $\sim 200$ at longer wavelengths, were fitted to each chunk,
using the IRAF CONTINUUM/ECHELLE procedure. 
These local continua were
then combined to produce a single continuum for the entire spectrum,
and any small discontinuities at the boundaries were adjusted manually
to produce a smooth result. 
We will call the continuum obtained in this way
the {\it initial} continuum, $C_{i}$.

As we will describe in the next section we have performed a full
joined Voigt profile analysis of the \ion{H}{i} and metal absorption
in the spectra in order to obtain absorption line parameters (the
redshift $z$, the column density $N$ in cm$^{-2}$ and the Doppler
parameter or $b$ parameter in \kms\/). The Voigt profile analysis is
sensitive to the assumed continuum and the simultaneous fitting of
different transitions caused by the same ion often reveals where the
continuum should lie in absorbed regions of the spectrum.  We have
therefore adjusted our initial continuum $C_{i}$ when we fitted
absorption features with the Voigt profile fitting routine
VPFIT{\footnote{Carswell et al.:
http://www.ast.cam.ac.uk/$\sim$rfc/vpfit.html}}.  The fitting
procedure and the adjustment of the continuum level are very similar
to the one described in Carswell, Schaye \& Kim (2002) and will be
discussed in more detail in Kim et al. (2007, in preparation).

For each spectrum, we first searched for metal absorption using the
entire range of the available spectrum. The identified metal lines
were the first to be fitted. During this process, the continuum was
adjusted to obtain acceptable ion ratios. If the metal lines were
blended with \ion{H}{i} absorption features, the \ion{H}{i} and metal
lines were fitted simultaneously.  Once all identified metal
absorption was fitted we then fitted the rest of the absorption
features assuming they are due to \ion{H}{i} absorption. For this we
used higher-order Lyman series lines whenever they were available to
constrain saturated lines.  In doing so, we further adjusted the
continuum level to achieve a satisfactory fit for the available Lyman
series. Using this {\it second} continuum estimated from the {\it
first} profile fitting procedure, the profile fitting was repeated and
the second continuum was checked manually again. This procedure was
repeated several times, until we were satisfied with the results.  The
fitting was performed both by R. F. Carswell and T.-S. Kim
independently, and the final fitting was carried out by T.-S. Kim.
Note that we did not account for the possibility of an extended
smoothly varying component of weak absorption often referred to as
Gunn-Peterson absorption. We further assumed that the response
function of UVES is smoothly varying for the non-calibrated spectra as
seen in a smoothly varying master response function provided by ESO.

The {\it final} adjusted continuum obtained from the fitting procedure
was then used to obtain the final normalised spectra used in this
study.  When we use the term ``continuum'' in the following, it always
refers to this {\it final} continuum, $C_{f}$. The difference between
$C_{f}$ and $C_{i}$ is small and usually restricted to some limited
wavelength regions, except in the regions around Lyman limit systems
and sub-damped Ly$\alpha$ systems where the difference becomes rather
large. We stress once again that these
final continua are {\it different} from the continua used in
\citet{kim04} which were estimated using the
$C_{i}$ procedure only. Note also that these final continua are not necessarily
always completely smooth since in some regions of the spectrum
continuum adjustments have been applied using a straight line to get
reasonable column density ratios for different transitions. These
small adjustments are required mainly because of the characteristics
of the un-calibrated spectra, especially at shorter wavelengths at
$\le$ 3400 \AA\/.  The number of adjustments depends on the QSO and
varies from none to several. Using a comparison between calibrated and
non-calibrated spectra, we were able to correct some non-smooth
features caused by instrumental artifacts.

\cite{kirk05} have performed a very detailed study of the localised
continuum fitting procedure used here using 24 high resolution, high
S/N Keck/HIRES spectra.  They tested their continuum fitting
uncertainties using real and artificial spectra fitted by 4 different
people.  \cite{kirk05} found noticeable differences between the fits
by different people. Similarly, R. F. Carswell and T.-S. Kim did not
agree on some parts of continuum fits of the spectra, confirming that
such a continuum fitting procedure is not fully objective.  However,
given the complexity of the problem we cannot see a better way of
doing it.  We will later quantify the effect of continuum
uncertainties on the flux probability distribution function and the
effective optical depth due to \ion{H}{i} absorption in
Section~\ref{sec4} and \ref{sec5}.
  
\begin{figure*}
\includegraphics[width=18cm]{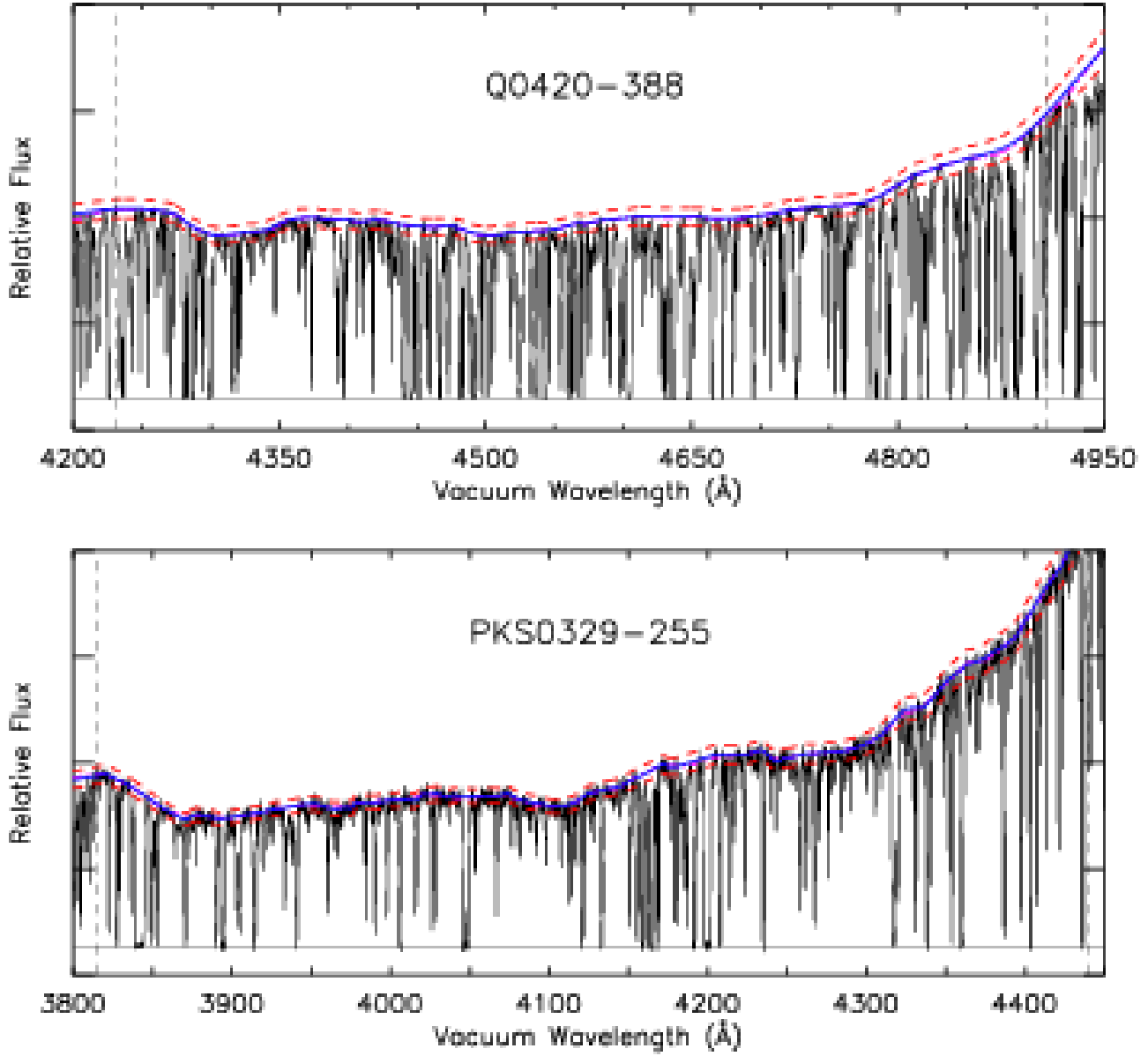}
\caption{{\protect\footnotesize{One of the best 
(Q0420$-$388 with $z_{\rm em}=$3.166, upper panel) and one of the worst
(PKS0329$-$255 with $z_{\rm em}=$2.704, lower panel) quality spectra
in our sample. The solid curves show our final continuum fit, while the
dashed curves show a continuum level increased/decreased by 5\%. 
The dot-dash curves which are almost indistinguishable
from the solid curves are the initial continuum fit, $C_{i}$. The
metal absorption has not been removed in both spectra.  The spectrum
of Q0420$-$388 is flux-calibrated and has one of the highest
signal-to-noise ratios. We estimate the uncertainty of its continuum
fit to be less than 1\%.  The spectrum of PKS0329$-$255 is not
flux-calibrated and has  low signal-to-noise (S/N $\sim 30$--40) 
in most parts of the spectrum.  
It also
shows the largest residuals from the fitted line profiles and the
largest zero flux offset. We estimate the continuum uncertainty to be
3--4\% in the lower signal-to-noise regions. 
The vertical thin-dashed lines indicate the minimum and
maximum wavelengths used in the analysis.}}}
\label{fig2}
\end{figure*}

The continuum uncertainty depends strongly on the S/N of the spectrum
as listed in Table~\ref{tab1}. The higher the S/N is, the smaller the
continuum uncertainty is.  In the upper panel of Fig.~\ref{fig2} we
show the calibrated spectrum of Q0420$-$388, one of our best spectra
with a S/N$\,=\,$100--140. The lower panel shows the non-calibrated
spectrum of PKS0329$-$255, one of the two worst quality spectra in our
sample.  For illustrative purpose we show the effect of a 5\% decrease
($C_{-5}$) and increase ($C_{5}$) of the continuum level by the dashed
curves.  It is obvious that $C_{5}$ and $C_{-5}$ dramatically
over/under-estimate the continuum level in the case of
Q0420$-$388. Even for PKS0329$-$255, $C_{5}$ and $C_{-5}$
over/under-estimate the continuum in most parts. Unfortunately, a rigorous
quantitative assessment of the continuum uncertainty is not possible
given the complex nature of the continuum fitting process.

In Table~\ref{tab1} we give rough estimates of the continuum
uncertainty obtained from absorption-free regions of the spectra.  
For this we compared  the highest  and the lowest flux levels 
in  absorption-free regions to the average flux level  in the same
regions, disregarding localised excursions in only one pixel. Note that
the continuum  uncertainty is not constant across the spectrum.  
Note further that due to the possible presence of extended featureless
absorption, the Gunn-Peterson absorption, the continuum used in this study
is more likely to be an underestimate than an overestimate of the true 
continuum.

The two lowest quality spectra in the sample are PKS0329$-$255
(lower panel of Fig.~\ref{fig2}) and J2233$-$606.  At 3700 \AA\/,
J2233$-$606 has a S/N of 30--35, and for PKS0329$-$255 the S/N is
generally $\sim 30$--40. The continuum uncertainties are $\sim$ 3--4\%
for the lower S/N regions and $\sim$ 2--3\% elsewhere.  PKS0329$-$255
also has the largest zero flux-level offset (see the next subsection)
in our sample.  PKS1448$-$232 and Q0122$-$380 also have continuum
uncertainties of $\sim 3$--4\% in regions where the S/N is
$\sim$30--35, at $\le 3550$ \AA\/ and at $\le 3450$ \AA\/,
respectively.
Fortunately most our spectra
have higher S/N, and the continuum uncertainties are generally around
1--2\%. 

\begin{figure*}
\vspace{-2.0cm}
\includegraphics[width=18cm]{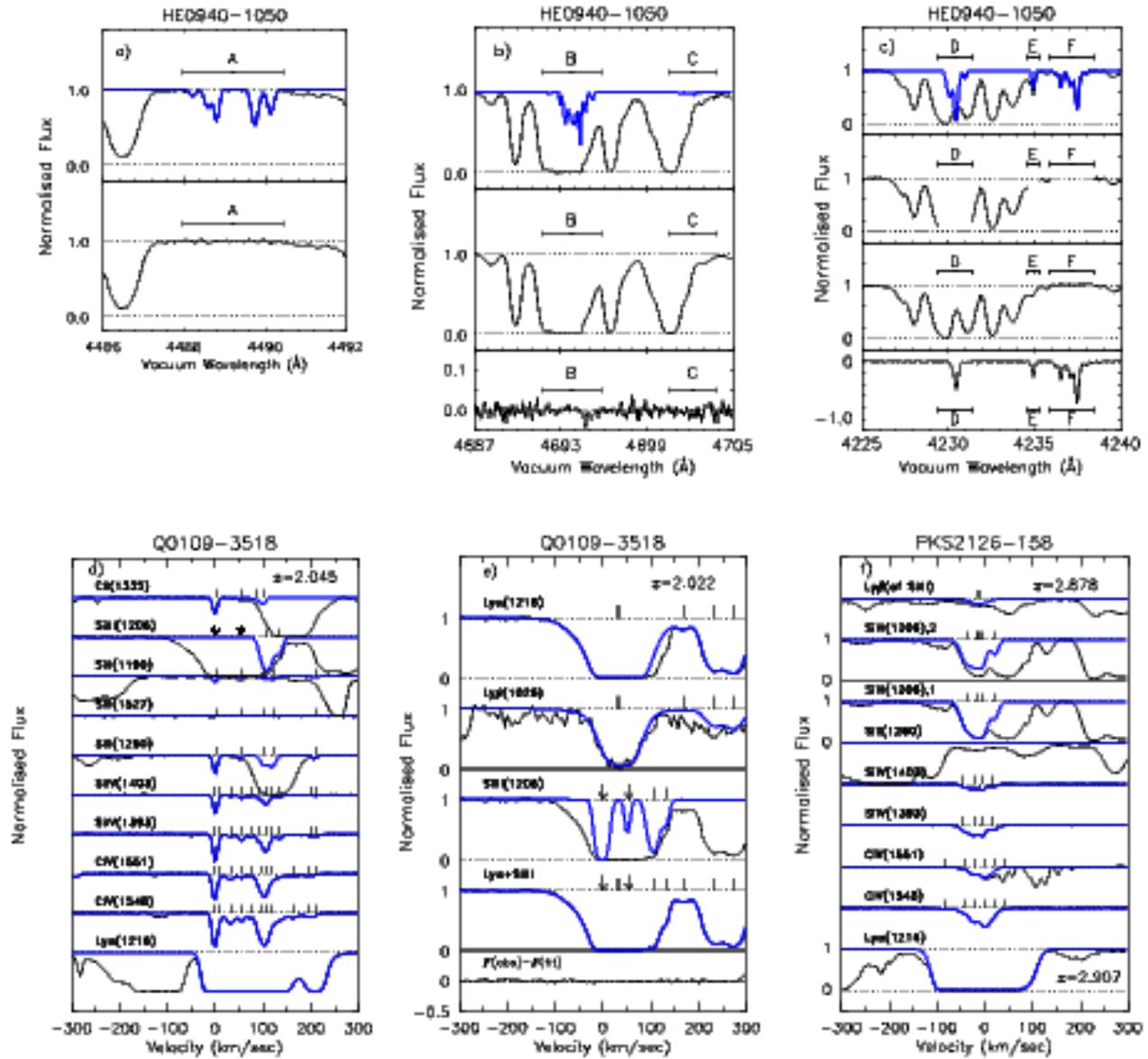}
\caption{{\protect\footnotesize{Various examples of metal absorption 
in the \lya forest region of the spectra in our sample.  Thin and
thick curves represent observed spectra  and fitted line profiles,
respectively. 
{\bf a)} Region A in the upper panel shows an isolated metal absorption
system, \ion{Si}{iv} 1393 at $z=2.221$ towards HE0940$-$1050.
\ion{Fe}{ii} 1608 at $z=1.789$ at 4485.4--4487.2 \AA\/ is not shown
in order to make the figure simpler.
The lower panel shows the spectrum after subtracting the metal
absorption in region A and adding noise from nearby absorption-free
regions.
{\bf b)} The thick solid curve in the top panel shows \ion{Fe}{ii}
1608 at $z=1.919$ (region B) and \ion{Fe}{ii} 1611 at $z=1.919$
(region C) in the spectrum of HE0940$-$1050. The line parameters for
both absorption systems have been obtained from other less strongly
blended \ion{Fe}{ii} transitions at other wavelengths.  The fitted
\ion{H}{i}-only profile is shown in the middle and the difference
between the observed profile and the fitted \ion{H}{i} profile is
shown in the bottom panel. 
{\bf c)} Regions D, E and F in the top panel mark absorption by
\ion{C}{iv} 1548 at $z=1.732$, \ion{C}{iv} 1548 at $z=1.735$ and a
mixture of \ion{C}{ii} 1334 at $z=2.175$ and \ion{C}{iv} 1550 at
$z=1.732$ in the spectrum of HE0940$-$1050.  The second panel shows
the regions of this part of the spectrum not contaminated by metals.
The fitted \ion{H}{i}-only profile and the difference between the
fitted \ion{H}{i}-only and the observed profile are shown in the third
and fourth panel, respectively.
{\bf d)} The absorption profile of a metal absorption system at
$z=2.045$ in the spectrum of Q0109$-$3518. The wavelength scale 
in this and the next two panels has been transformed to a corresponding
Doppler velocity.  The profiles of the \ion{Si}{iv} and \ion{Si}{ii}
absorption suggest that two \ion{Si}{iii} 1206 components exist at 0
\kms and $\sim 54$ \kms   (indicated by the arrows) which are blended
with saturated \ion{H}{i} absorption.
{\bf e)} The velocity profile for \ion{Si}{iii} 1206 in the same system as
in {\bf d)}.
The top panel shows the \ion{Si}{iii} region at $z=2.045$. The
superimposed line is the \ion{H}{i} Ly$\alpha$ profile at
$z=2.022$. The second panel is  the \ion{H}{i} Ly$\beta$ profile at
$z=2.022$. The profile fit used both Ly$\alpha$ and Ly$\beta$
simultaneously, assuming no \ion{Si}{iii} (at $z=2.045$) contribution.
The third panel shows the  \ion{Si}{iii} absorption profile
with the two components at 0 and $\sim 54$ \kms, assuming the
same line strength ratio for these two components as 
estimated from the resolved \ion{Si}{ii} and \ion{Si}{iii} absorption.  The
fourth panel is the generated profile of the \ion{H}{i} ($z=2.022$, top panel)
and \ion{Si}{iii} absorption ($z=2.045$).
The bottom panel is the difference between the observed and the
generated \ion{H}{i} $+$ \ion{Si}{iii} profiles.
{\bf f)} The velocity profiles of a metal line system at $z=2.907$
towards PKS2126$-$158. The second panel shows the
\ion{Si}{iii} profile assuming the contribution from the superimposed
\ion{H}{i} at $z=2.878$ which was estimated from the corresponding
Ly$\beta$ profile (top panel).  The third panel shows the  
\ion{Si}{iii} profile assuming no contribution from \ion{H}{i}.  
The  weak, broad  \ion{Si}{ii} 1260 absorption (the strongest
\ion{Si}{ii} transition in the optical spectra) in the fourth panel is
assumed to be a non-detection.}}}
\label{fig3}
\end{figure*}

\subsection{Uncertainty of the zero flux level}

As discussed in \cite{kim04}, the UVES standard pipeline reduction
program returns spectra where the sky level is somewhat
under-subtracted.  This problem becomes more severe at shorter
wavelengths ($\le 3400$ \AA\/). The under-subtraction of the sky level 
results in a non-zero flux level within the troughs of saturated
lines. The offset from zero flux depends on signal-to-noise and
wavelength. For increasing signal-to-noise and longer wavelengths the
offset from zero becomes smaller.  The worst case is PKS0329$-$255
with a typical offset, $\Delta F_{\rm zero} =0.015$ in a spectrum
normalised to unit continuum. Note that this QSO also has one of the worst
overall continuum uncertainty. Most spectra have an offset of much less
than $\Delta F_{\rm zero} =0.01$. An offset from zero flux can lead to
significant problems for the Voigt profile fitting. Since the flux in
saturated regions does not go to zero, VPFIT normally adds many narrow
components whose flux minima do not reach $F \sim 0$ instead of
fitting the saturated profiles with fewer strong components, unless
one specifically allows the zero level to vary.  Fortunately, this
problem has a negligible effect on the statistics of the flux
distribution presented here, as will be discussed in more detail in
Section~\ref{sec4:4}.

\section{A full  Voigt profile analysis of the \ion{H}{i} and metal absorption}
\label{sec3}

Most simulations and models of the \lya forest account only for the
distribution of \ion{H}{i}. In reality, the \lya\ forest  contains
significant absorption from intervening metals in the IGM. These metal
absorption lines are obviously interesting in their own right and
provide a wealth of information on the spectrum of the metagalactic UV
background and the metal enrichment of the IGM by galaxies. For a
statistical analysis of the \ion{H}{i} absorption aimed at studying
the underlying matter distribution, they are, however, a source of
contamination which leads to additional uncertainties.  In this case
it is therefore important to identify the metal lines in the forest
and to remove them from further study, or at least to quantify their
effect on any statistic investigated.  Most studies so far have dealt
with this problem by excluding spectral regions with strong metal
contamination from the analysis, e.g. a strategy adopted by
\citet{rau97}, \citet{mc00} and \citet{lid06}.

The metal contamination is, however, rather widespread and varies
strongly between different spectra. Excluding contaminated regions is
therefore problematic.  If metal lines are isolated, it is
straightforward to excise the metal-contaminated regions.  More often
than not, metal absorption is, however, blended with \ion{H}{i}
absorption at $2<z<3.5${\footnote{The amount of the metal
absorption blended with \ion{H}{i} absorption depends on redshift. At
low redshift ($z < 2.5$) the \ion{H}{i} absorption is 
weaker and most metal absorption occurs in region with weak or no
\ion{H}{i} absorption. At higher redshift ($z > 2.5)$, the number
density of \ion{H}{i} absorption features increases and the metal
absorption starts to blend in regularly with the \ion{H}{i}
absorption. At very high redshift ($z > 3.5$--4), the density of
\ion{H}{i} absorption features becomes very high and many lines become
saturated. In this case, many metal lines are blended with strongly
saturated \ion{H}{i} absorption and their contribution to the total
absorption diminishes or becomes negligible.}}.
It is then difficult to decide which regions to exclude.  Since there
are only a few QSOs in each redshift bin in this and other similar
studies, attempts to exclude metal contaminated region are likely to
lead to a selection bias, especially at $z \sim 3$.

We take a different approach here, capitalising on a full
Voigt-profile analysis of the \ion{H}{i} and metal absorption
performed by R. F. Carswell and T.-S. Kim which will be described in
more detail elsewhere (Kim et al. 2007, in preparation). In this Voigt profile
analysis we have identified all metal lines in the 18 spectra to the
best of our knowledge. This allows us to remove their absorption
contribution from the observed flux distribution.  This method
produces an estimate of the continuous flux distribution without any
absorption due to {\it identified} metal lines.

For the identification  of metal lines, we made use of  known
properties of metal absorption: 
\begin{enumerate}

\item Metal absorption features  tend to
be narrower  than \ion{H}{i} lines.

\item Metal absorption features 
are usually associated with strong, often saturated \ion{H}{i} absorption.

\item In the redshift range 
considered here, it is unlikely for \ion{H}{i} absorption 
to show corresponding \ion{Si}{ii} without showing 
\ion{C}{iv}, \ion{C}{ii}, \ion{Si}{iv}
or \ion{Si}{iii}. In short, \ion{H}{i} absorption with  no
corresponding \ion{C}{iv} is unlikely to show other metal lines,
especially for the weaker absorption systems with column density 
$N_{\ion{H}{i}} \le 10^{14} \, \mathrm{cm}^{-2}$.

\end{enumerate}

We started by identifying metal absorption features redward of the
Ly$\alpha$ emission where all absorption lines (apart from the
telluric lines, which were omitted) are due to metals.  Most of these
are \ion{C}{iv}, \ion{Si}{iv}, \ion{Si}{ii}, \ion{Al}{ii},
\ion{Al}{iii}, \ion{Mg}{ii} and \ion{Fe}{ii}. The rest-frame
wavelengths and oscillator strengths for transitions
from these ions are given by \citet{morton2003}.
Using these lines as a
guide, we looked for other metal lines at the same redshift, such as
\ion{C}{iii}, \ion{O}{vi}, \ion{C}{i}.  In a second step we searched
for metal lines associated with \ion{H}{i} 
absorption with any column
densities in the UVES spectra, i.e. at $\ge 3050$ \AA\/.  
Thirdly we looked for possible doublet candidates in the
forest. In a fourth and final step we looked for possible metal lines
associated with saturated \ion{H}{i} absorption or with the Lyman
discontinuity shown in the HST data listed in Table~\ref{tab1},
i.e. at $\le 3050$ \AA\/.

With information on absorption systems at $z < 1.5$ available for only
half of the QSOs (see Table~\ref{tab1}) and none of them observed at
as high a resolution and S/N as the UVES data, it is very difficult to
identify {\it all} the metal lines in the forest.  Severe line
blending often makes the identification of metal lines
difficult. Sometimes an absorption feature first thought to be due to
\ion{H}{i} did not show a corresponding \lyb\/ line of the expected
strength when we attempted to fit higher-order Lyman series lines.
Any excess absorption at the \lya\/ wavelength was then considered as
being due to yet to be identified metal lines.
 
All other absorption not identified as being due to metals we have
assumed to be due to \ion{H}{i}. There were a few complexes of strong
clustered rather narrow lines (Doppler parameters less than 15
\kms) which we suspect to be unidentified metal lines. We have flagged
these as potential metal lines. These occasions, however, were rare.
Note that some of the weak narrow lines which we considered as due to
\ion{H}{i} absorption could still be either weak, unidentified metal
lines or noise peaks, while stronger metal lines are almost complete
in their identification.

The different metal lines and \ion{H}{i} were all fitted
independently. Only the transitions produced by the {\it same ion} of
the same redshift systems were required to have the same redshift and
the Doppler parameters.  Sometimes part of a given metal line is
blended with strong \ion{H}{i} absorption which makes it difficult to
separate the metal and \ion{H}{i} lines.  {\it Only} in such
circumstances, the redshifts and the Doppler parameters were {\it
tied} with the ones measured from the other clean metal lines within
the same ionisation group (such as \ion{Mg}{ii}, \ion{Fe}{ii},
\ion{C}{ii} and \ion{Si}{ii} for the low-ionisation group), similar to
the method usually employed in the metal analysis of damped \lya
systems.  Fortunately, most common metal lines embedded in the forest,
such as \ion{Mg}{ii}, \ion{Fe}{ii}, \ion{Si}{ii}, \ion{C}{iv}, are
multiplets.  Identifying these lines is more reliable than
single-transition lines, such as \ion{Si}{iii} 1206.

While we are likely to have identified almost all strong metal lines,
we would like to stress again that we will inevitably have ascribed
some weak lines wrongly as being due to \ion{H}{i}.

Fig.~\ref{fig3} shows some examples of the metal lines embedded in the
forest. It also illustrates how we have subtracted the metal
contributions from the flux distribution.  Thin lines are the observed
profiles (\ion{H}{i}+metal lines), while the thick lines are the
{\it fitted} \ion{H}{i} or metal lines. 
The upper panel of Fig.~\ref{fig3} a) shows
a single isolated metal system and the lower panel shows the same
region after removing the metal lines and adding the noise estimated
from nearby, absorption-free regions.  About 40$-$50\% of metals are
isolated at $z \sim 2$, the fraction decreases to 20$-$30\% at $z \sim
3$, due to increased Lyman line blending.

The top panel of Fig.~\ref{fig3} b) shows a moderate-strength metal
line blended with \ion{H}{i} absorption. Such embedded metal lines
have been identified by other transitions falling outside the forest
region or in regions of the forest where the \ion{H}{i} absorption is
weak.  The middle panel shows the \ion{H}{i}-only absorption profile
generated from the parameters obtained from the line fitting.  The
lower panel shows the difference between the flux in the top 
(\ion{H}{i}+metal) and
middle panels. The difference is of the same order as the pixel
noise. If metal absorption of weak to moderate strength is blended
with saturated \ion{H}{i} absorption, the metal absorption contributes
very little to the combined absorption profile.

Fig.~\ref{fig3} c) shows a very common configuration of metal
absorption.  Approximately $45$\% of metal absorption systems at $z
\sim 2$ and $\sim 75$\% at $z \sim 3$ look similar to this. The top
panel shows a region of a spectrum overlaid with the metal absorption
profile of identified metal lines. In region D the metal absorption is
blended with moderate-strength \ion{H}{i} absorption. If we wanted to
excise the metal absorption it is not obvious if one should cut out
the entire \ion{H}{i} absorption complex from 4226 \AA\/ to 4236 \AA\/
or only the region where the metals affect the \ion{H}{i} profile from
4229.5~\AA\/ to 4231.2~\AA\, as shown in the second panel.  The third
panel shows instead the \ion{H}{i}-only absorption profile generated
from the parameters obtained from our line fitting procedure.  The
bottom panel shows the difference between the \ion{H}{i} and the
\ion{H}{i}+metal profiles.

The most difficult metal absorption features to obtain a reliable line
fit for in the Ly$\alpha$ forest region of the spectrum is the
absorption due to the single transition of \ion{Si}{iii} at
1206.5~\AA\/ (see McDonald et al. 2006). Sometimes we could
successfully fit \ion{Si}{iii} absorption profiles blended with
\ion{H}{i} absorption using other metal lines, such as Ly$\alpha$, Ly$\beta$,
\ion{Si}{ii}, \ion{Si}{iv}, \ion{Al}{iii}, \ion{Al}{ii}, \ion{C}{ii},
\ion{C}{iv}, and \ion{Mg}{ii}.  Sometimes, however, we could fit only
part of a \ion{Si}{iii} profile.  In such cases some
contamination of the final \ion{H}{i} profile is unavoidable, since we
could not subtract all of the \ion{Si}{iii} contribution from the
\ion{H}{i}+\ion{Si}{iii} absorption feature.
An example is shown in Fig.~\ref{fig3} d) and e).

Fig.~\ref{fig3} d) shows a typical metal line system at $z=2.045$
towards Q0109$-$3518 where \ion{Si}{iii} is blended with \ion{H}{i}
absorption at $z=2.022$.  Note that we plot relative velocities here
instead of wavelength as is customary for strong metal absorption
complexes.  The low (high) ionisation lines \ion{Si}{ii}
(\ion{Si}{iv}) suggest that there should be two components of
\ion{Si}{iii} at 0 \kms and $\sim 54$ \kms   in the saturated \ion{H}{i}
profile at $z=2.022$, as indicated by the arrows.  The top panel of
Fig.~\ref{fig3} e) shows the \ion{Si}{iii} region of the system shown
in panel d).  Superimposed is the \ion{H}{i} Ly$\alpha$ profile at
$z=2.022$, which was fitted together with the corresponding Ly$\beta$
absorption (second panel), assuming no \ion{Si}{iii} (at $z=2.045$) 
contribution.
The third panel shows the \ion{Si}{iii}
profile with the two components at $v = 0$ and $\sim$ 54 \kms
(indicated by the arrows).  These two components were generated
assuming the same relative line strengths between all {\it
successfully} fitted \ion{Si}{ii} and \ion{Si}{iii} components.  In
the 4th panel the \ion{H}{i} Ly$\alpha$ (at $z=2.022$) $+$
\ion{Si}{iii} components (at $z=2.045$) are superimposed.  Note that
the fit parameters of the \ion{H}{i} profile are the same as in the
first panel.  The bottom panel shows the difference between the
observed and the fitted \ion{H}{i} Ly$\alpha +$~\ion{Si}{iii}
profiles.  The figure demonstrates that if the \ion{H}{i} absorption
is saturated, then even strong metal lines do {\it not} contribute
significantly to the overall profile, unless the metal absorption
extends to the wings of the saturated \ion{H}{i} absorption profiles.

Fig.~\ref{fig3} f) shows a velocity plot of a metal line system at
$z=2.907$ towards PKS2126$-$158.  It illustrates the uncertainty of the
profile fit for a single transition \ion{Si}{iii} 1206 and a very weak
line.  The second and the third panels show the \ion{Si}{iii} 1206 absorption,
assuming a \ion{H}{i} contribution at $z=2.878$ constrained by the
corresponding Ly$\beta$ (the top panel indicated by a thick tick
mark) and assuming no \ion{H}{i} contribution, respectively.  In
reality, its {\it true} column density could range between the ones
obtained with these two assumptions. Similarly, the weak, broad feature at the
expected location for \ion{Si}{ii} 1260 (the strongest \ion{Si}{ii}
transition available in the spectra) in the fourth panel could be a
real detection blended with weak \ion{H}{i} (the upper limit for
$N_{\mathrm{\ion{Si}{ii}}}$ can be estimated from non-detection of 
other \ion{Si}{ii} transitions above the Ly$\alpha$ emission)
or a weak broad \ion{H}{i}. Since we do not want to
over-identify metals and want to be {\it conservative}, we adopted  the
lowest \ion{Si}{iii} 1206 column density (assuming a contribution by
\ion{H}{i}) and considered  \ion{Si}{ii} 1260 to be not detected in this case.  Only
when it was not possible to constrain \ion{H}{i} from higher order
Lyman series, we assumed that the  absorption was  due to metals
only. Non-detection in Ly$\beta$ in our  spectra typically
corresponds to $N_{\mathrm{\ion{H}{i}}} \le 10^{12.8}$~cm$^{-2}$
assuming $b=20$~\kms\/ (or $F \ge 0.8$ in Ly$\alpha$). 
Fortunately
there are not many such cases in our sample. In Table~\ref{tab_b} in
Appendix A we list uncertain line fits.

In summary, for each QSO we have obtained a Ly$\alpha$ forest spectrum 
free of identified metal lines as follows:
\begin{itemize}
\item for isolated metal lines the spectrum was replaced by continuum
with a noise level estimated from nearby continuum regions;  
\item where metal lines and \ion{H}{i} lines are blended, that part of
the spectrum was replaced with the model \ion{H}{i} lines determined
from fitting Voigt profiles to the heavy element and Lyman lines 
simultaneously, with noise using estimates appropriate to the final
flux levels determined from nearby regions.
\end{itemize}

In Fig.~\ref{fig4} we show the continuum fitted spectra of our
complete sample together with the metal contribution to the total
absorption to give a general impression of the metal
contamination. Note that the metal contribution to the total
absorption differs significantly from the metal-only absorption as
metal absorption blended with strong \ion{H}{i} has little effect on
the total absorption profile.

\begin{figure*}
\vspace{-0.1cm}
\includegraphics[width=16cm]{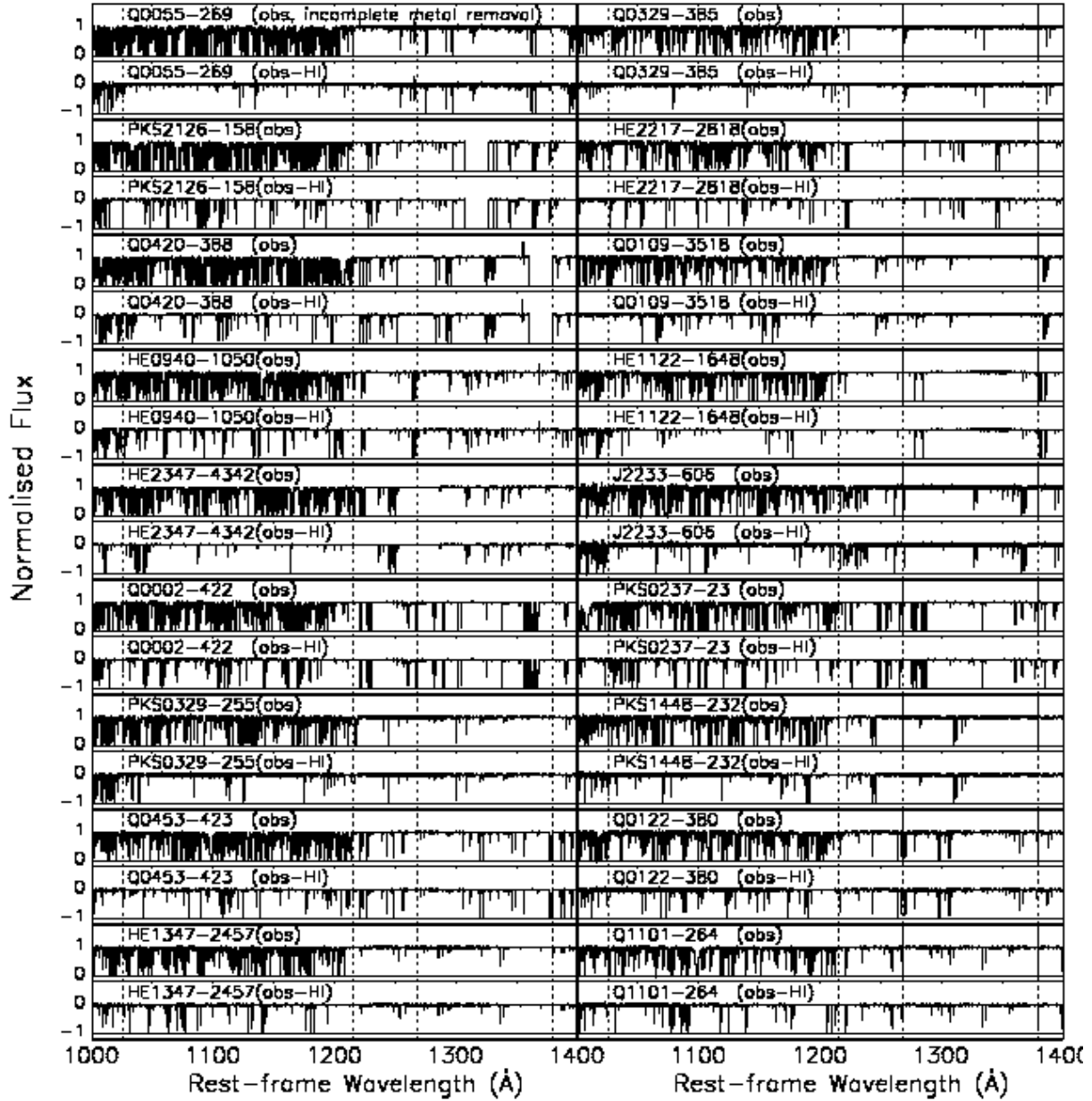}
\caption{{\protect\footnotesize{Normalised observed spectra and the
additional metal contribution to the absorption for our sample of 18
VLT/UVES spectra (against rest-frame wavelength). The regions with
no flux, such as at $\sim 1315$~\AA\/ of PKS2126$-$158, 
are due to the wavelength gaps caused by using three separate
CCDs in the dichroic setting.  The normalised observed spectra are
shown by the thick solid curves in the upper panels for each QSO. Note
that the lower panel is not the fitted metal absorption. It is the
difference between the observed flux and the fitted \ion{H}{i}
absorption and thus shows the additional absorption by metals which is
significantly smaller than the absorption by metals would be in the
absence of \ion{H}{i} absorption. The four vertical dotted lines
indicate the rest-frame wavelength of Ly$\beta$, Ly$\alpha$ plus
1268~\AA\/ and 1380~\AA\/. Note that the weak metal absorption at
1268--1380~\AA\/ does not necessarily indicate a low metal
contamination in the forest
(see e.g. the spectra of HE0940$-$1050
and HE1347$-$2457). The most common metal species found in high-$z$ QSO spectra
is \ion{C}{iv}. The high column density absorption systems showing 
strong \ion{C}{iv} (1548~\AA\/
and 1550~\AA\/) and \ion{Si}{iv} (1393~\AA\/ and 1402~\AA\/) absorption features 
usually show other
metal lines.
These high column density absorption systems occur randomly along each
sightline, i.e. at different $z$ for
different sightlines. Therefore, 
stronger transitions e.g. \ion{Fe}{ii} 2382 \& 2600, \ion{Si}{ii} 1190,
1193 \& 1260, or \ion{Mg}{ii} 2796 \& 2803, which are usually associated with
high column density systems, could be present in the Ly$\alpha$ forest
region without there being any corresponding strong metal lines in the
1268--1380~\AA\/ range.
The metal contribution in the spectrum of
Q0055$-$269 forest could be significantly underestimated since the
wavelength coverage in the red is rather limited, only up to 6809~\AA\/
(observed). 
}}}
\label{fig4}
\end{figure*}

\section{Statistical errors and systematic Uncertainties of the flux probability distribution}
\label{sec4}

\subsection{Statistical errors}
\label{sec4:1}

One of the main aims of the paper is to estimate the \ion{H}{i}-only
absorption flux PDF from the normalised spectra
of our sample.  The PDF of the flux is simply the number of pixels
which have a flux between $F$ and $F + \Delta F$ for a given flux $F$
divided by the total number of pixels \citep{jen91, rau97, bryan99,
mc00, kim01}.

Before presenting our new measurement of the flux PDF from our full sample we
will discuss the statistical errors of the measured flux PDF  
and investigate some of the systematic uncertainties discussed 
in Section~\ref{sec2} on the flux.  Following \cite{mc00},
we calculate the flux PDF for bins of width $\Delta F =0.05$.
Pixel with flux level smaller than $F=0.025$ or greater than $F=0.975$
have been allocated to the $F=0$ bin and the $F=1.0$ bin, respectively. 

To estimate the errors of the PDF both for a single QSO and for a full
sample, we used a modified  jackknife method as presented by
\cite{lid06}.   We briefly outline the method here for clarity.  We
first calculate the flux PDF of the full sample.
This sample was then divided into  $n_{c}$ chunks with a
length of $\sim$50 \AA\/.    If the PDF estimated
from the full sample at the flux bin $F_i$ is $\widehat{P}(F_i)$ and 
the PDF estimated without the $k$-th 
chunk at the flux bin $F_i$ is $\widetilde{P}_{k}(F_i)$, then 
the covariance matrix cov$(i,j)$ between the PDF in a flux bin $F_i$
and the PDF in a flux bin $F_j$ was calculated as 

\begin{equation}
\mathrm{cov}(i,j) = \sum_{k=1}^{n_{c}} [\widehat{P}(F_i)-\widetilde{P}_{k}(F_i)]
[\widehat{P}(F_j)-\widetilde{P}_{k}(F_j)],
\end{equation}
 
\noindent and the variance at a given flux level is given by the
diagonal terms of the covariance matrix $\sigma_{i}^2 =
\mathrm{cov}(i,i)$ for a flux bin $F_i$.  We checked that this
modified jackknife method is not sensitive to the length/number of
chunks for a given sample size. As expected, the errors are larger
when the number of pixels in the sample is smaller: $\sim$15\% for
Q0420$-$388 with 13561 pixels vs $\sim$7\% for the full $<\!z>\,=2.94$
sample with 34265 pixels at $F=0.5$. We compared the errors obtained
with the modified jackknife method with the errors obtained by 500
bootstrap realisations of chunks of 100 pixels (or 5~\AA\/) used by
\cite{mc00} and \cite{sch03} (see Section~\ref{sec5:2} for more
details for the bootstrap method).  Both methods give comparable
error estimates, while the Poisson errors (i.e. those based on the
square root of the number of pixels) tend to be $\sim 35$\% smaller.

\subsection{Continuum fitting}
\label{sec4:2}

We now move to a discussion of the effect of systematic uncertainties on
the flux PDF.  We start with the effect of continuum fitting
uncertainties.  In Fig.~\ref{fig5} we show the flux probability
distribution of the spectrum of Q0420$-$388 including the metal
absorption (see also Fig.~\ref{fig6}) with our final continuum fit
$C_{f}$, our initial continuum fit $C_{i}$ 
and four further continua where we have applied a wavelength
independent offset of the continuum level of $\pm 1\%$ ($C_{1}$ and
$C_{-1}$) and $\pm 5\%$ ($C_{5}$ and $C_{-5}$). The dotted, dashed,
solid, dot-dot-dot-dashed and dot-dashed curves show the PDF with
$C_{5}$, $C_{-5}$, $C_{f}$, $C_{1}$ and $C_{-1}$, respectively.
The flux PDF of the spectra with the initial continuum fit $C_{i}$ is
almost indistinguishable from the PDF of the spectra with the final
continuum fit $C_{f}$. 
As expected changing the continuum level affects the PDF most strongly at
a flux level of $F\sim 1$, shifting the corresponding peak in the
PDF. Continuum fitting uncertainties also have a moderate effect on
the slope of the PDF at flux levels $0.65 < F < 1.0$.  There is little
effect at lower flux levels as regions of saturated or very strong
absorption are not affected by an {\it over-/under-}estimated continuum
as much as regions of weak absorption.
 
Note again that a systematic change of the continuum level by 5\% is a
gross overestimate of the actual continuum fitting uncertainty for
most of our spectra. We choose this value here simply to demonstrate the effect
more clearly.  Most spectra in our sample have a continuum uncertainty
of $\sim 1$--2\%, and the changes to the PDF are small.

\begin{figure}
\includegraphics[width=9cm]{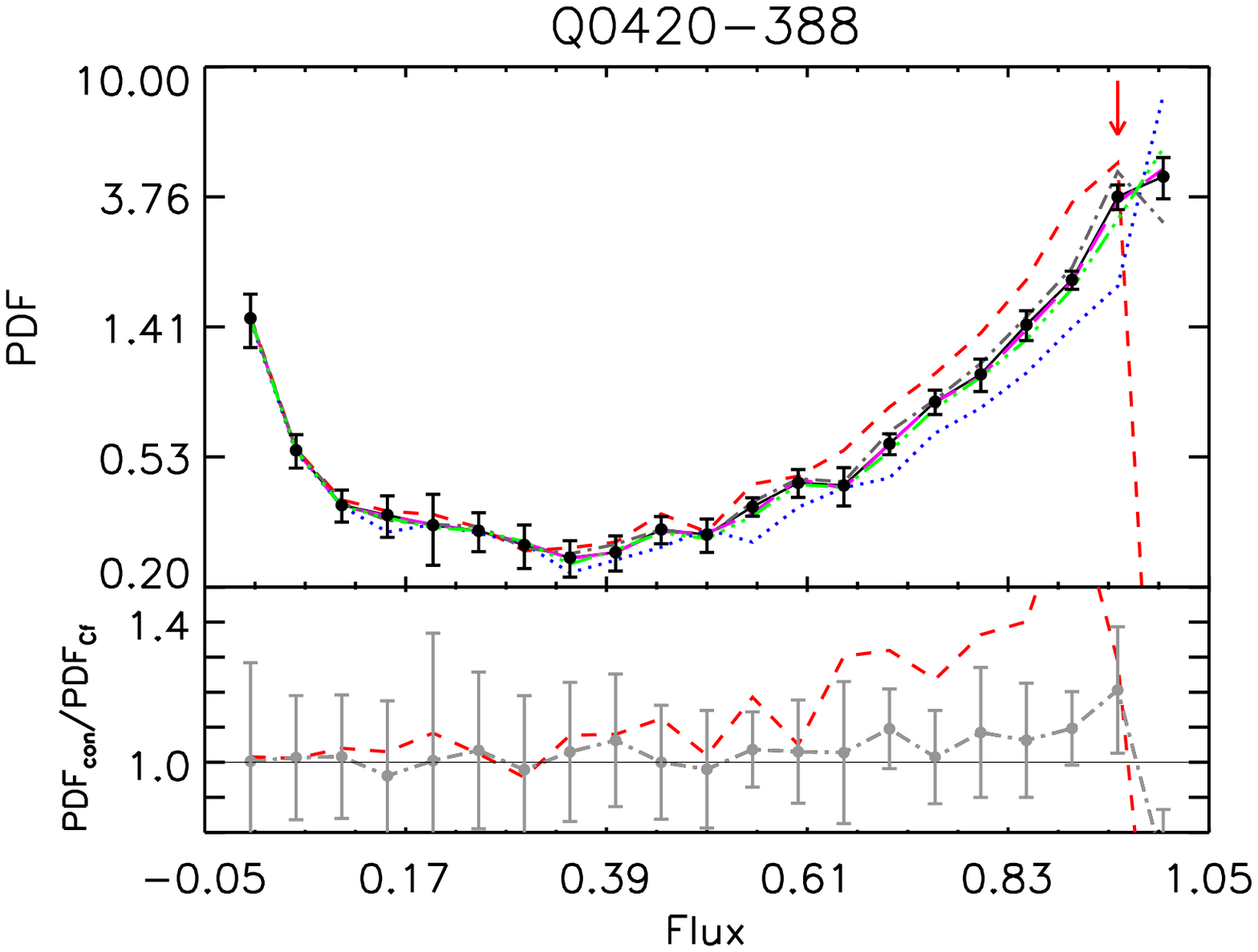}
\vspace{-0.5cm}
\caption{{\protect\footnotesize{The upper panel shows the flux PDF of
Q0420$-$388 ($2.480 < z < 3.038$) for 6 different continuum levels.
The solid curve is the PDF for our best fit final continuum $C_{f}$, 
while the curve almost indistinguishable from the solid curve is the PDF
for the initial continuum $C_{i}$. The
dotted/dashed curve are the PDF for a continuum level increased/decreased by
5\% ($C_{5}$ and $C_{-5}$), while the dot-dot-dot-dashed/dot-dashed curve
are for a continuum level increased/decreased by 1\% ($C_{1}$
and $C_{-1}$) The errors were estimated using a modified jackknife method
by Lidz et al. (2006) as described in the text.  Different continuum
fits also shift the peak of the PDF, as indicated by the arrow for
$C_{-5}$.  The lower panel shows the following ratios
PDF$_{{C}_{-5}}$/PDF$_{{C}_{f}}$ (dashed curve) and
PDF$_{{C}_{-1}}$/PDF$_{{C}_{f}}$ (dotted curve). The errors in the
lower panel are similar for both curves but plotted only once for
clarity. The errors correspond to the combined errors of both
PDF$_{{C}_{-5}}$ and PDF$_{{C}_{f}}$.
}}}
\label{fig5}
\end{figure}

\subsection{Metal contamination}
\label{sec4:3}

As apparent from Fig.~\ref{fig4} the metal contribution to the
absorption varies significantly between different spectra.  In
Table~\ref{tab_d} in Appendix A we quantify the metal contamination
of the \ion{H}{i}+metals effective optical depth,
i.e. observed values before removal of the metal
absorption, estimated using  
the effective optical depth after removal of the metal absorption.
The values vary from
0.5\% to 28\% (note that in the case of Q0055$-$269, the metal removal
is incomplete due to the limited wavelength coverage from
3200$-$6810 \AA\/). 
There appears to be
no obvious trend of the extent of the metal contamination with other
parameters, such as $z$, except that spectra containing Lyman limit
systems or sub-damped Ly$\alpha$ systems tend to show a larger
absorption contribution by metals. The mean metal contamination is
$\sim$12\%. This is consistent within the  errors quoted for {\it
statistical} estimates of the metal contamination: $\sim 19$\% at
$z=1.9$ by \cite{tyt04} and $\sim 10$\% at $z=2.7$ by
\cite{kirk05}{\footnote{Both removed the metals in the forest
statistically.  Using published line parameters {\it redwards} of the
Ly$\alpha$ emission of QSOs at $1.7 < z_{\mathrm{em}} < 3.54$, they
estimated the amount of metals as a function of rest-frame and
observed wavelength.}}.
  
Fig.~\ref{fig6} illustrates the effect of absorption by metals on the
flux PDF. The upper panel shows the PDF of two artificial spectra
which were generated from the fitted line parameters of
PKS2126$-$158. The solid curve is the PDF after removal  of the identified metal
lines  while the dotted curve is the PDF before removal of the metal
absorption.  The bottom panel shows the
ratio of the two.  Unlike continuum uncertainties metal absorption
affects the flux PDF mostly at flux levels in the range $0 < F < 0.6$
even though there is also a small reduction of the flux PDF at $F\sim
1$. Note that the latter effect is likely to be underestimated as weak
metal absorption is difficult to identify.  Since the number of
absorption features which we classified as {\it suspected} metal lines
are small and usually weak, these lines have a negligible effect on
the PDF at $0.1 < F < 0.6$.

\begin{figure}
\includegraphics[width=9cm]{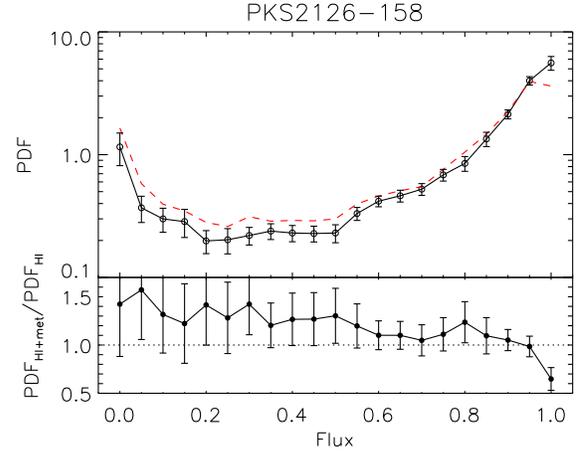}
\caption{{\protect\footnotesize{
The upper panel shows the flux PDF of PKS2126$-$158 with metal
absorption (dashed curve) and without metal absorption (solid curve). The
two spectra were generated using  fitted line parameters. 
The lower panel shows the ratio  of the flux  PDF with and without metals.
The errors in the lower panel are the combined errors.
The metal lines contribute to the flux PDF significantly at flux
levels $F < 0.6$. }}}
\label{fig6}
\end{figure}

\subsection{Pixel noise and the uncertainty of the zero flux level}
\label{sec4:4}

Pixel noise will cause a slight smoothing of the flux PDF. For
high-S/N spectra as in our sample the effect is small but noticeable.
This is demonstrated in Fig.~\ref{fig7}. The upper panel shows the
flux PDF of PKS0329$-$255 for the observed spectrum with metals
(dot-dashed curve) and without metals (solid curve with error bars).
The observed spectrum has S/N $=30$--55.  We generated 4 artificial
spectra with different S/N using the fitted line parameters, assuming
Gaussian noise: the dashed curve is for S/N $=25$, the dotted curve is
for S/N $=50$, and the two almost indistinguishable thin solid curves
are for S/N $=100$ and S/N $=\infty${\footnote{Note that the noise in
the observed spectra is not exactly Gaussian.  The noise at $F\sim 0$,
i.e. the bottom of saturated lines, and at $F \sim 1$ are 
different, and of course the S/N at $F \sim 0$ is much
worse than at $F \sim 1$.  However, for the purpose of illustrating
the S/N effect on the PDF, the assumption of Gaussian noise is a good
approximation.}}.
Note that the generated spectra do not have a offset from the zero flux
level, i.e. the saturated lines go down to $F=0$.  The bottom panel
shows the ratio of the PDF for the artificial spectrum with
S/N $=\infty$ (solid curve) and the PDF for the spectrum with metals
(dot-dashed curve) to the PDF of the metal-removed spectrum.

The effect of the signal-to-noise on the flux PDF is most evident at
flux levels of $0 < F < 0.1$ as well as at $F > 0.9$, where it can exceed that
from the metal contamination. At intermediate flux levels, $0.1 < F <
0.9$, the effect of the signal-to-noise is comparable or smaller than
that due to metal absorption. It should, however, be noted here that
the contribution of metal absorption in the spectrum of PKS0329$-$255
is rather small (2.8\%) and that the spectrum has the lowest S/N of
our sample.

In Fig.~\ref{fig8} we show the difference between  the flux PDF
of artificial S/N = $\infty$ spectra (i.e. no zero-level offset) 
and the flux PDF of the observed 
spectra without metal absorption for our full sample at three
different redshift bins.  At flux levels of $0.1 < F < 0.8$ the
difference is less than 1\% and much smaller than the statistical
errors, but outside this range the errors increase to a few percent 
at low flux levels and more than 10\% at high flux levels. 

\begin{figure}
\includegraphics[width=9cm]{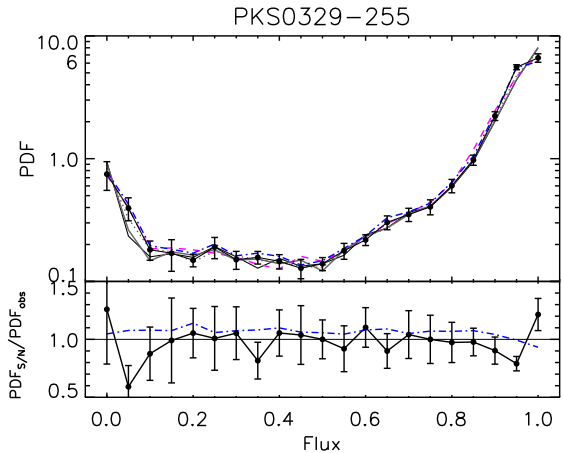}
\caption{{\protect\footnotesize{
The upper panel shows the flux PDFs of the observed and artificial spectra
of PKS0329-255 for different noise levels. The dot-dashed curve is for
the observed spectrum including the metals, the solid curve with
errors is for the observed spectrum after the metals have been
removed. The four other curves are the flux PDFs for artificial
spectra generated from the line parameters with different levels of
Gaussian noise added.  The dashed curve is for S/N $=25$, the dotted
curve is for S/N $=50$ and the two almost indistinguishable 
thin solid curves are
for S/N $=100$ and S/N $=\infty$.  The S/N $=\infty$ and S/N $=100$ PDFs
show a rather large deviation at $F < 0.1$ and almost a factor of 2 at
$F=0.05$.  This is due to the offset of the zero flux level of the
observed spectrum (a typical offset from zero flux is $\sim 0.015$)
which the artificial spectra do not have and the effect of which is
not smoothed out in the artificial spectra with the highest S/N. The
lower panel compares the ratio of the flux PDFs of 
the S/N $=\infty$ spectrum to the metal-removed spectrum (solid curve)
and the PDF ratio of the metal-included spectrum to the metal-removed
spectrum (dash-dotted curve).
The effect of low S/N on the flux PDF becomes noticeable at $F < 0.1$
and $F > 0.9$. At $0.1 < F < 0.9$, the effect of the S/N is
comparable or smaller than that of the metal contamination.
}}}
\label{fig7}
\end{figure}

\begin{figure}
\includegraphics[width=9cm]{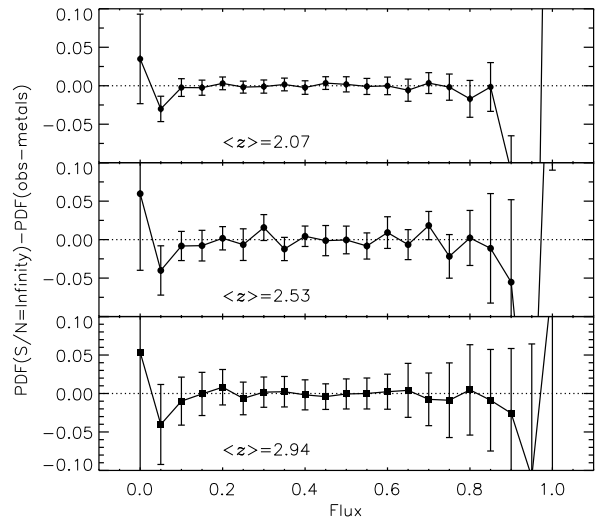}
\vspace{-1cm}
\caption{{\protect\footnotesize{
The difference between the flux PDFs for  S/N $=\infty$ artificial
spectra and metal-removed observed spectra for the full sample
divided into three redshift bins. The errors are those of the 
PDF of the metal-removed observed spectra. The combined 
errors of both PDFs are larger. The effect of lower S/N and 
the offset of the zero flux level on the PDF are  negligible 
compared to the statistical errors  at flux levels $0.1 < F < 0.8$.
}}}
\label{fig8}
\end{figure}

\section{The flux distribution of the full sample}
\label{sec5}

\subsection{The flux probability distribution}
\label{sec5:1}

Fig.~\ref{fig9} shows the main result of this paper, the flux PDF of
the observed spectra in our full sample divided into three redshift
bins (Table~\ref{tab_a} in Appendix A) after removal of all identified
metal lines. The thin, grey curves show the flux PDFs of the
individual spectra. As found previously there is considerable scatter
between different lines-of-sight. 
This is mainly due to the
occurrence of strong absorption systems which are rare in individual
spectra \citep{viel04b}. 
A considerable path length is therefore
required to reach reasonable convergence to an average flux PDF.
This is best illustrated by the spectrum of Q0002$-$422 which shows the most
deviant individual PDF compared with the mean PDF in its redshift bin.
This is probably due to several factors. The spectrum falls at the
upper end of its redshift bin. It has also the shortest usable path length
(191 \AA\/) and the largest number of strong systems per unit redshift
in the wavelength region used.

In Fig.~\ref{fig10} we show the effect of the removal of the
identified metal absorption. The dashed and solid curves in the upper
panel show the flux PDF of the spectra with and without the identified
metal absorption, respectively. The lower panel shows the ratio of the
two.  Removing the metal absorption mainly affects the flux PDF at
flux levels $0.1<F<0.8$.  Without metal absorption the fraction of
pixels in this flux range is 10$-$20\% lower.  The effect of metal
absorption, which appears to evolve little with redshift, is more
significant at lower redshift where the \ion{H}{i} absorption is
smaller.

In Fig.~\ref{fig11} we compare our new measurement of the flux PDF
with that of \cite{mc00} at $<\!z\!> \, = 2.41$ and $<\!z\!> \, =
3.0$.  The solid curves are the PDF when metals are included.
Unfortunately a comparison at higher redshifts is not possible due to
the lack of high-redshift spectra in our sample.  At $<\!z\!> \, =
2.41$ our measurement is about 10--20\%  lower at flux levels
$0.1<F<0.5$ while at $<\!z\!> \, = 3.0$ our measurement is about
10--30\%  
lower at flux levels $0.2<F<0.8$.  We can only speculate here
where this discrepancy comes from.  Part of the difference is probably
due to the rather crude removal of metal absorption by McDonald et
al., which is likely to have led to more residual absorption by
unidentified metal lines in their spectra.  The difference appears,
however, to be larger than expected due to this effect and increases
rather than decreases with increasing redshift. Note that the McDonald
et al.  sample is significantly smaller with a total of 8 spectra and
some of the discrepancy can probably be explained as being due to
variations between different lines-of sight (see the grey thin curves
in Fig.~\ref{fig9}).  Differences in the placement of the continuum
level may also play a role.  We list the mean flux PDF for the five
different redshift bins discussed in this section in Table~\ref{tab_c}
in Appendix A.

\begin{figure*}
\includegraphics[width=16cm]{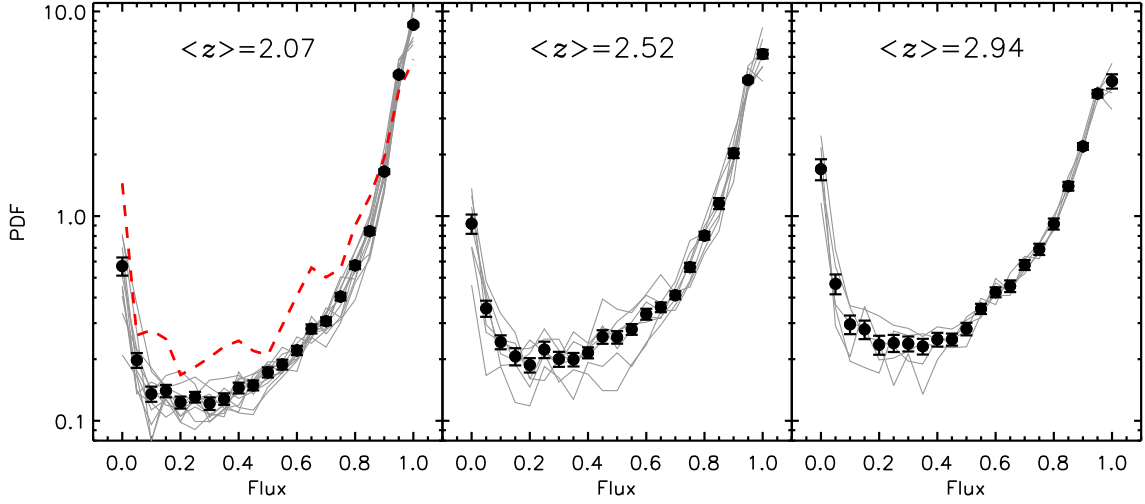}
\vspace{-6.2cm}
\caption{{\protect\footnotesize{The flux PDF (filled circles with errors) 
of the full sample
divided into three redshift bins after removal of the metal
absorption. The thin grey curves show the PDFs of individual spectra.
The 1$\sigma$ error bars are estimated using a modified jackknife method
(Lidz et al. 2006).  The dashed curve in the $<\!z\!> \, = 2.07$ bin
is the PDF of the spectrum of Q0002$-$422, which has the highest
redshift in this bin and the largest effective optical depth.}}}
\label{fig9}
\end{figure*}

\begin{figure*}
\includegraphics[width=16cm]{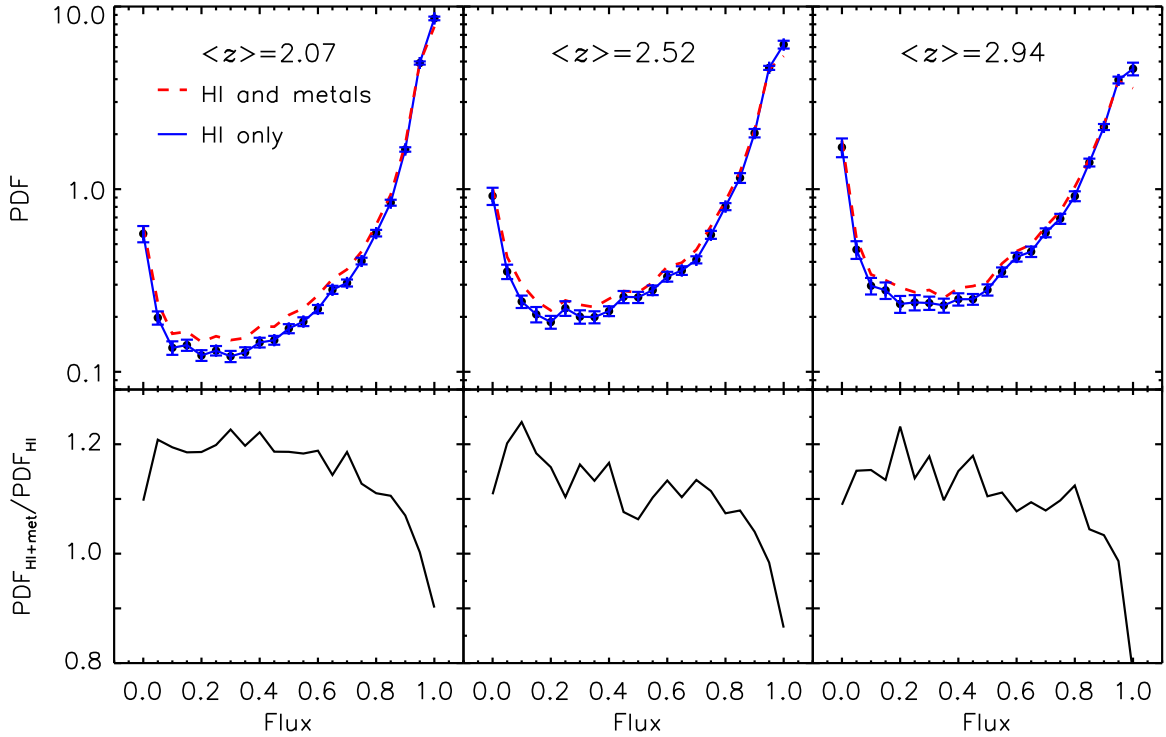}
\vspace{-3.2cm}
\caption{{\protect\footnotesize{The flux PDF of the full sample
divided into three redshift bins before (dashed curves) and after
(solid curves) removal of the metal absorption.  The lower panel
shows the ratio of the PDFs before and after removal of the metal
absorption. The effective optical depth due to \ion{H}{i} absorption is
lower at lower redshift and the relative contribution of the metal
absorption is therefore larger.}}}
\label{fig10}
\end{figure*}

\begin{figure*}
\includegraphics[width=16cm]{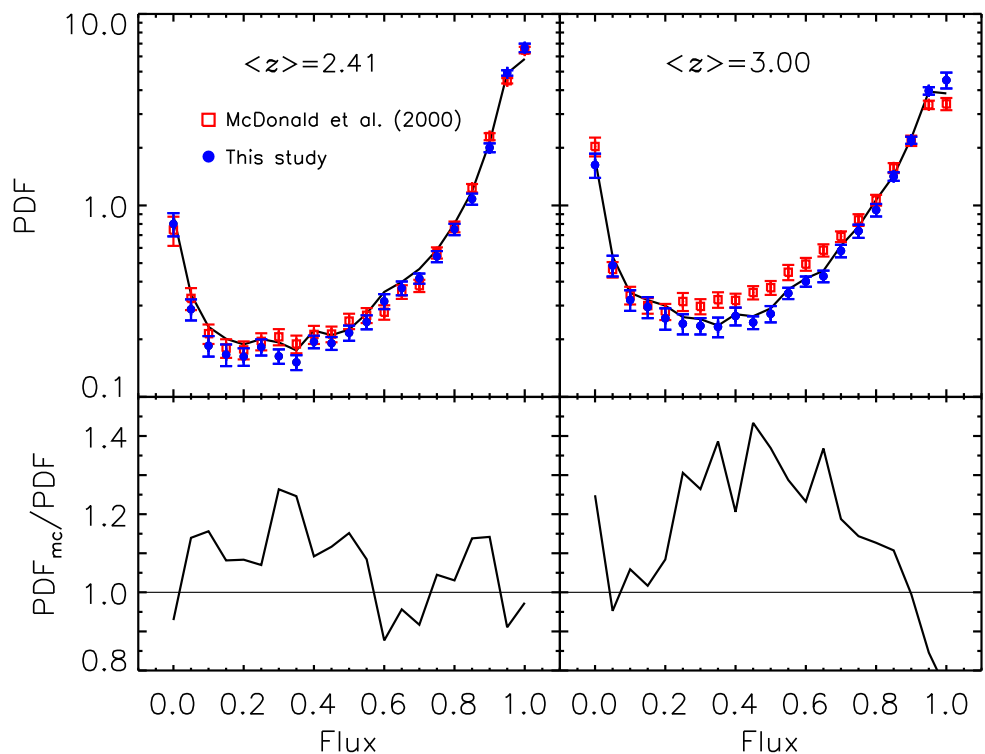}
\vspace{-3.2cm}
\caption{{\protect\footnotesize{Comparison of the flux PDF of our
sample divided into two redshift bins with that of McDonald et
al. (2000).  The solid curve and filled circles show the flux PDF of
our sample before and after removal of the metal absorption,
respectively.  The lower panel shows the ratio of the flux PDF of
McDonald et al. to that of our sample after removal of the metal
absorption.  }}}
\label{fig11}
\end{figure*}

The flux PDF is a statistical quantity which -- at least in principle
--  can be easily compared with that of simulated spectra. Due to the
expected tight correlation between \ion{H}{i} optical depth, local gas density
and temperature, the observed flux PDF can constrain astrophysical parameters
of the IGM as well as cosmological parameters \citep{weinberg98,
mc00, mek01,  des05}. The effect of the removal of metal absorption
and the difference to the published flux PDF by McDonald et al.
is comparable or larger than the effect of changing some of
the model  parameters within plausible ranges (e.g. Meiksin et
al. 2001; Desjacques \& Nusser 2005). Our new improved measurement of
the flux PDF should therefore be relevant for attempts to use the flux
PDF to constrain astrophysical and cosmological parameters.

\subsection{The \ion{H}{i} effective optical depth}
\label{sec5:2}

So far we have concentrated on the flux PDF of our sample. The
\ion{H}{i} effective optical depth 
($\tau_{\ion{H}{i}}^{\mathrm{eff}}$)  
is  another intensely studied quantity
which is important for the  comparison with  models of the \ion{H}{i} 
distribution.  The \ion{H}{i} effective optical depth is related to the mean
flux  as $\exp^{-\tau_{\ion{H}{i}}^{\mathrm{eff}}} = \ <\! \exp^{-\tau_{\ion{H}{i}}}\!>$ ,
where $<\,\,>$ indicates the mean value averaged over wavelength. 
Note that the effective optical depth is {\it not} the average of the optical depth.
The effective optical depth is the quantity which directly goes into
measurements of the amplitude of the
metagalactic UV background  \citep{rau97,bolton05}
and plays a crucial role in calibrating measurements of the matter
power spectrum from \lya forest data  based  on the flux power
spectrum  \citep{cro02, sel03, tyt04, viel04, lid06}.
 
As we have taken special care with the removal of the metal absorption 
in our spectra it is worthwhile to revisit the effect the 
removal of the metal absorption has on the \ion{H}{i} effective optical
depth.

\begin{figure*}
\includegraphics[width=16cm]{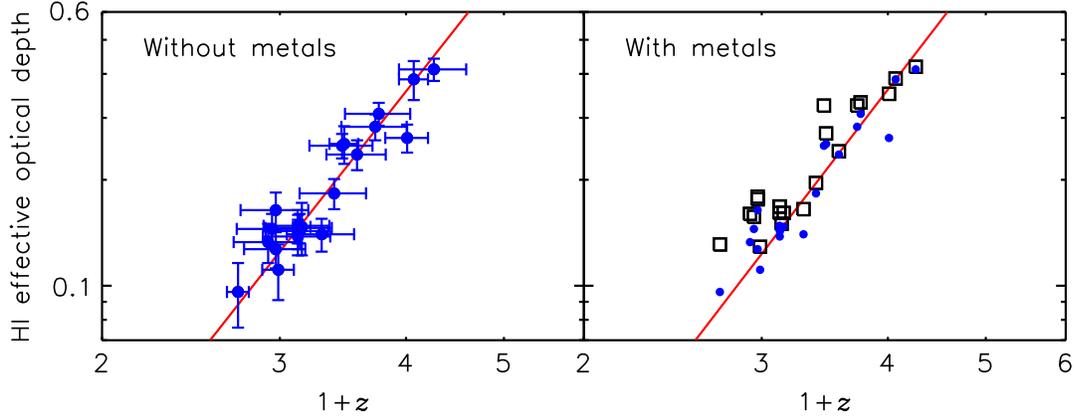}
\vspace{-7.5cm}
\caption{{\protect\footnotesize{Left panel: Evolution of the effective
optical depth of our full sample after removal of the metal
absorption. 
The horizontal and vertical error
bars show the redshift range used and the $1\sigma$ errors.
The $1\sigma$ errors were estimated as in McDonald et al. (2000) and
Schaye et al. (2003) using 500 bootstrap realisations of chunks of 100
pixels (5 \AA\/).
Right panel: Comparison of the evolution of the
effective optical depth of our full sample before (open squares) and
after (filled circles) removal of the metal absorption.  The solid
lines in both panels represent a power-law fit to the effective
optical depth $\tau_{\ion{H}{i}}^{\mathrm{eff}}$ 
of our sample after removal of the metal absorption
and other measurements from high-resolution data compiled from
the literature at $1.7 < z < 4$ as in
Fig.~\ref{fig13},
$\tau_\ion{H}{i}^{\rm eff} = 
(0.0023 \pm 0.0007) (1+z)^{3.65 \pm 0.21}$.
}}}
\label{fig12}
\end{figure*}
 
In Fig.~\ref{fig12} we show the effective optical depth of our
observed spectra before (open squares) and after (filled circles) 
removal of the
metal absorption.  The solid lines in both panels is the 
power-law fit of $\tau_{\ion{H}{i}}^{\mathrm{eff}}$ to the 
optical depth of the high-resolution data from our sample
and that compiled from the literature,
$\tau_\ion{H}{i}^{\rm eff} = 
(0.0023 \pm 0.0007) (1+z)^{3.65 \pm 0.21}$
(the same power-law fit as in Fig.~\ref{fig13}).
The errors of \eff were estimated with the
same procedure adopted in \cite{mc00} and
\cite{sch03}{\footnote{{Note that the modified jackknife method
using a $\sim 50$ \AA\/-long chunk generally gives error estimates of
only $\sim$1--2\% for each QSO. These estimates are too low, since a $\pm
1$\% continuum uncertainty generally gives an error of $\sim 2$--10\%.}}.
Each spectrum was divided into chunks of 100 pixels (or 5 \AA\/). We
then performed 500 bootstrap realisations, treating each chunk as a
one data point. The removal of the metal absorption leads to a
typical reduction in the {\it observed} effective optical depth
(i.e. \ion{H}{i}+metals) by 0.5\% to 28\%
with the mean of $\sim 12$\%, albeit as expected with a large
scatter (see Table~\ref{tab_d} in Appendix A).  As mentioned in
Section~\ref{sec4:3}, these numbers are estimated from the effective
optical depths of the observed spectra before and after removal of
the metal absorption.
The \ion{H}{i} effective
optical depth is obviously also subject to the same systematic
uncertainties as the flux probability distribution. The change of
continuum from $C_{f}$ to $C_{5}$ ($C_{1}$) increases \eff\/ by
0.049 (0.010), which is identical for each QSO. 
The noise and the uncertainty in the
zero-level flux have a negligible affect on \eff\/.

The upper panel of Fig.~\ref{fig13} compares the $\tau_\ion{H}{i}^{\rm
eff}$ measurements of our 18 QSOs with those of Schaye et
al. (2003). Both set of values are measured from spectra where metal
absorption has been removed.  In the case of Schaye et al. (2003) this
has been done by excising spectral regions contaminated with strong
metal absorption.  The values are in good agreement. Considering that
half of the sample of spectra studied here is part of the sample of
\cite{sch03} this is not too surprising.  For spectra common to both
studies the differences are, in most cases, less than 10\% when our
wavelength ranges were adjusted to the ones used in the Schaye et al. 
sample{\footnote{There are several noticeable differences between the two
\eff\/ measurements. Since the two \eff values of the metal-included
spectra are very similar, the difference is mainly caused by the
incomplete metal removal in the Schaye et al. sample, i.e. their \eff\/
is somewhat  larger.  QSOs with more than 10\% difference in \eff are:
J2233$-$606 at $1.732 < z < 1.963$ by $\sim$12\%, Q0109$-$3518 at
$1.874 < z < 2.120$ by $\sim$11\% and Q0002$-$422 at $2.434 < z < 2.710$
by $\sim$11\%.
In Fig.~\ref{fig13}, the most noticeable is J2233$-$606,
in part due to the fact that our redshift range used in this study
differs from the one used by Schaye et al.}.
Table~\ref{tab_d} in Appendix A lists the measurement of \eff\/ as
well as the mean flux, its variance and the contribution of metal
absorption in the forest.  The errors are estimated from the 500
bootstrap realisations of a chunk of 5 \AA\/.

In the lower panel of Fig.~\ref{fig13} we have compiled a range of
\eff\/ measurements from the literature at $1.7 < z < 4$
and compare this with a
measurement of $\tau_\ion{H}{i}^{\rm eff}$ from our sample after
removal of the metal absorption binned in redshift bins with width
$\Delta z = 0.2$ (filled circles with errors). 
Our \eff\/ is estimated from the mean flux of all
pixels in each bin, instead of averaging the effective opacities of
each QSO (see Table~\ref{tab_e} in Appendix A}).  The errors are
estimated by the 500 bootstrap realisations of 5 \AA\/ using all pixels
in each bin.  The open squares, filled squares and filled triangles show
the measurements of \cite{sch03}, \cite{kirk05} and \cite{mc00},
respectively. The filled diamond at $z=1.86$ and the open circle at
$z=1.9$ are the measurements of \cite{jan06} and \cite{tyt04},
respectively. Note that our values are in good agreement
with those measured by \cite{kirk05} at $1.7 < z < 3.3$, 
although the latter 
did not include any LLSs and removed the metal contribution statistically.

The solid line in Fig.~\ref{fig13} is the best power-law fit to 
the optical
depth values of high-resolution data in the redshift range $1.7 < z <
4$ both from the literature as shown in the figure 
and our sample presented here. The errors in the \eff\/ measurements
are taken into account in the fit. Despite
the different method of measuring \eff\/, we also included 
the Kirkman et al. measurement.
Note that the
point at $z=1.9$ by \cite{tyt04} is from low resolution data and is
therefore not included in the fit.  We did not include the
\cite{sch03} measurements from the spectra which are already included
in our sample. The fit is given by $\tau_{\ion{H}{i}}^{\rm eff} =
(0.0023 \pm 0.0007) (1+z)^{3.65 \pm 0.21}$.
For our sample alone the power law is $\tau_\ion{H}{i}^{\rm eff} =
(0.0054 \pm 0.0101) (1+z)^{2.96 \pm 0.83}$. This fit is shallower  
than that of the combined \eff\/ due to a small number of QSOs and a lack of
high-$z$ QSOs in our sample
(cf. Kim et al. 2001; Kim et al. 2002).
The dashed line is an
extrapolated fit by \cite{ber03} to their low resolution, low
signal-to-noise SDSS data (their S/N $> 4$ sample)
in the redshift range $2.5 < z < 4$, while
the dot-dot-dot-dashed line is a fit by \cite{fan06} to measurements
at $3 < z < 5.5$ extrapolated to lower redshift,
$\tau_{\ion{H}{i}}^{\rm eff} \propto (1+z)^{4.3 \pm 0.3}$. The slope
of the power law evolution measured by Bernardi et al. is in 
good agreement within 1$\sigma$. 
As discussed extensively in the
literature (e.g. Seljak et al. 2003; Viel et al. 2004b) there is, 
however, a systematic
offset of about $\sim 25$--30\% between the Bernardi et al. \eff\/ and
our and other measurements from high-resolution, high S/N data. This is
most likely attributable to the difficulty of continuum fitting for
low resolution, low signal-to-noise data which appears to lead to a
systematic overestimate of the effective optical depth.  The Fan et
al. measurement (the dot-dot-dot-dashed line) has a somewhat steeper
redshift evolution. This might indicate that there is a deviation from
a single power law at $z > 4$.  Note that \cite{becker07} have also
argued that the redshift evolution at $z >4$ steepens towards higher
redshift and is not well fit by a power law anymore. At $1.7 < z < 4$,
we find no evidence for a deviation from a power
law. Table~\ref{tab_e} in Appendix A lists the measurement of \eff\/
sampled at $\Delta z=0.2$ as well as \eff\/ of each QSO belonging to
each redshift bin.

\begin{figure}
\includegraphics[width=9cm]{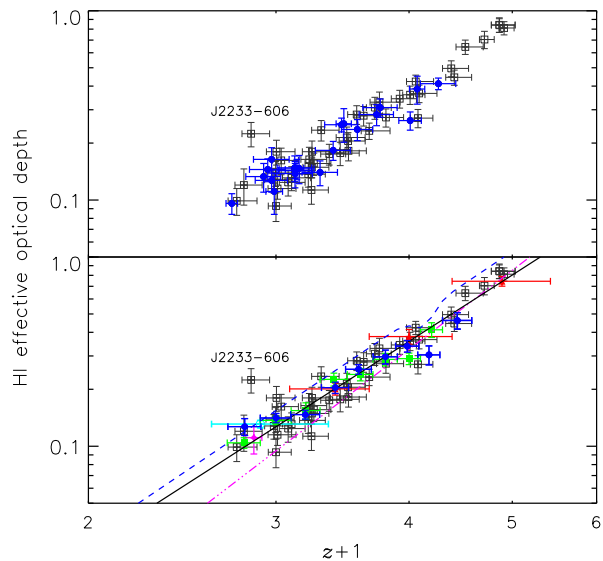}
\caption{{\protect\footnotesize{Comparison of the evolution of the
effective optical depth of our full sample after removal of the metal
absorption (filled circles) with that measured in Schaye et al. (2003,
grey open squares).  There is a large overlap in the two samples and
as expected the estimates from the two studies are in good agreement
(for J2233$-$606, see text). The horizontal and vertical error
bars show the redshift range used and the $1\sigma$ errors. The errors
were estimated using 500 bootstrap realisations of chunks of spectra
with 100 pixels.  The lower panel shows a comparison of the evolution
of the effective optical depth of our sample divided into redshift
bins with width $\Delta z=0.2$ (filled circles with errors)
to a compilation of other measurements
from the literature.  The open squares are the measurements of Schaye
et al. (2003) as in the upper panel.  The filled squares, the filled
triangles, the filled diamond at $z=1.86$, and the open circle at
$z=1.9$ are from Kirkman et al. (2005), McDonald et al. (2000),
Janknecht et al. (2006) and Tytler et al. (2004), respectively.  The
solid line is the best fit to all high-resolution data in the figure
both from the literature and the current sample
in the redshift range $1.7 < z < 4$: 
$\tau_{\ion{H}{i}} = (0.0023 \pm 0.0007) (1+z)^{3.65 \pm 0.21}$.
The dashed curve is a fit from
Bernardi et al. (2003) to their S/N $> 4$ SDSS data 
in the redshift range from $2.5 <
z < 4$ extrapolated to lower redshift.
The dot-dot-dot-dashed line is a fit by Fan et al. (2006) to data in
the redshift range $3 < z < 5.5$ extrapolated to lower redshift.
}}}
\label{fig13}
\end{figure}

\section{Conclusions}
\label{sec6}

We have obtained improved measurements of the flux probability
distribution at $1.7 < z < 3.2$ and effective optical depth due to the
intergalactic \ion{H}{i} absorption at $1.7 < z < 4$, based on a
sample of 18 high resolution, high signal-to-noise VLT/UVES QSO
absorption spectra for which we have performed an extensive Voigt
profile analysis of the \ion{H}{i} and metal absorption.

The main results are as follows:

\begin{enumerate} 

\item{The normalised flux probability distribution (PDF) is 
      affected mainly by metals at the level of 10--20\% at flux
      levels of $0.2 < F < 0.8$ depending on redshift, and  
      by continuum fitting uncertainties at the level of 5--20\% 
      at flux levels of $0.8 < F < 1$, depending on 
      the signal-to-noise of the spectra. 
      The effects 
      of pixel noise and zero-level offset are very small, only
      noticeable at $F\sim 0$ and $F\sim 1$.  
      The metal contribution to the absorption varies from a 
      few percent to up to 30 percent between different lines of
      sight. A careful individual removal of the metal absorption is
      therefore essential for an accurate determination of the shape 
      of the PDF.}

\item{Our new measurements of the flux PDF due to \ion{H}{i} alone 
      are systematically lower at $0.2 < F < 0.7$ than the widely-used PDFs
      measured by \cite{mc00} based on a sample of Keck/HIRES spectra 
      half the size of our sample. The difference has the same sign as
      would be expected if the removal of the metal absorption 
      by \cite{mc00} was less complete than ours, but it is actually somewhat 
      larger and does not have the correct redshift dependence 
      if this were the sole cause. Given the rather small sample size
      cosmic variance is likely to be responsible for  
      part of the discrepancy.}
    
\item{The effect of our improved removal of metal absorption 
      on the measured effective optical depth due to \ion{H}{i}
      absorption, compared to previous measurements where the metal absorption 
      has been taken into account in simpler ways, is small. 
      Our new measurements of $\tau_{\ion{H}{i}}^{\rm eff}$
      are in good agreement with other measurements from 
      high resolution, high signal-to-noise spectra.  In the redshift
      range   $1.7 < z < 4$  the redshift  evolution of the 
      of our measurements of the effective optical depth 
      and other measurements from high resolution spectra 
      is  best fit with a single power law 
      $\tau_{\ion{H}{i}}^{\mathrm{eff}}
      = (0.0023 \pm 0.0007) (1+z)^{3.65 \pm 0.21}$.}
\end{enumerate}

\section*{Acknowledgments.} 
We would like to thank ESO, the ESO staff, the ESO science
verification team and the UVES LP team of ``The cosmic evolution of
the intergalactic medium'' for initiating, compiling and making
publicly available a superb set of QSO absorption spectra.
We also thank the UVES team for building the spectrograph
and our referee Michael Strauss for his useful suggestions.
TSK would
like to thank Michael Murphy for providing his spectral combining
program, UVES\_popler, and the IoA, Cambridge, UK, for hospitality during
the final stages of this work.

\newpage
\appendix
\label{appen}

\section[h]{Tables}

\begin{table}
\caption{The wavelength range of each spectrum in  the three redshift bins}
\label{tab_a}
\begin{tabular}{lccc}
\hline
\noalign{\smallskip}
QSO & $<\!z\!> \, = 2.07$  &  $<\!z\!> \, = 2.52$  &  $<\!z\!> \, = 2.94$  \\
\noalign{\smallskip}
\hline
\noalign{\smallskip}
Q0055--269          &            &            & 4785--5112 \\
PKS2126--158        &            &            & 4638--5112 \\
Q0420--388          &            & 4231--4506 & 4506--4909 \\
HE0940--1050        &            & 4197--4502 & 4502--4870 \\
HE2347--4342        &            & 4055--4502 & 4502--4643 \\
Q0002--422          & 3901--4092 & 4092--4504 & \\
PKS0329--255        & 3815--4090 & 4090--4439 & \\
Q0453--423          &            & 4084--4362 & \\
HE1347--2457        & 3705--4091 & 4091--4319 & \\
Q0329--385          & 3528--4105 &            & \\
HE2217--2818        & 3509--4091 &            & \\
Q0109--3518         & 3532--4070 &            & \\
HE1122--1648        & 3514--4082 &            & \\
J2233--606          & 3335--3886 &            & \\
PKS0237--23         & 3361--3865 &            & \\
PKS1448--232        & 3306--3860 &            & \\
Q0122--380          & 3306--3819 &            & \\
Q1101--264          & 3503--3765 &            & \\
\noalign{\smallskip}
\hline
\end{tabular}
\end{table}

\begin{table*}
\caption{Uncertain line fits for individual QSOs$^{\mathrm{a}}$}
\label{tab_b}
\begin{tabular}{ll}
\hline
\noalign{\smallskip}
QSO & Uncertain line fits \\
\noalign{\smallskip}
\hline
\noalign{\smallskip}
Q0055--269          &
        \ion{Si}{iii} at $z=3.2562$ (m),
        \ion{Si}{ii} 1260 at $z=3.2562$ (l) \\
PKS2126--158        & \ion{Si}{ii} 1260 at $z=2.9071$ (l),
      \ion{Si}{iii} at $z=2.9067$ (m),
      \ion{Si}{iii} at $z=2.7668$ (m)  \\
Q0420--388    & \ion{Si}{iii} 1260 at $z=2.5372$ (l),
       \ion{Si}{iv} 1394 at $z=2.2475$ (l),
       \ion{Si}{iv} 1403 at $z=2.2475$ (l)\\
HE0940--1050        & \ion{Si}{ii} 1260 at $z=2.8346$ (l),
         \ion{C}{ii} 1334 at $z=2.3306$ (l),
         \ion{O}{i} 1302 at $z=2.3292$ (l),
         \ion{C}{ii} 1334 at $z=2.2212$ (m) \\
HE2347--4342  &  \ion{Si}{iii} at $z=2.7412$ (m),
           \ion{Si}{ii} 1260 at $z=2.3124$ (l) \\
Q0002--422          & \ion{Si}{ii} 1260 at $z=2.4623$ (l),
           \ion{Si}{iii} at $z=2.4642$ (m),
           \ion{Si}{iii} at $z=2.3022$ (m) \\
PKS0329--255        & \ion{Si}{iii} at $z=2.3289$ (m) \\
Q0453--423          & \ion{Al}{ii} at $z=1.6172$ (l) \\
HE1347--2457  &
     \ion{C}{iv} 1548 at $z=1.6083$ (m) \\
Q0329--385          &  \\
HE2217--2818        &
           \ion{Si}{iii} at $z=1.9653$ (l), \ion{C}{ii} 1334
           at $z=1.690983$ (l) \\
Q0109--3518         & \ion{C}{ii} 1334 at $z=2.0463$ (m), \ion{Si}{iii} at
         $z=2.0473$ (l) \\
HE1122--1648        & \ion{Si}{iii} at $z=2.0071$ (m) \\
J2233--606          & \ion{Si}{iii} at $z=1.9426$ (h),
      \ion{Si}{iii} at $z=1.9257$ (m)
      \ion{Si}{iii} at $z=1.8711$ (m),
      \ion{C}{ii} 1334 at $z=1.8711$ (l) \\
PKS0237--23         & \ion{C}{iv} 1548, 1550 at $1.3643$ (m),
           \ion{Si}{iii} at $z=2.1781$,
           \ion{C}{ii} 1334 at $z=1.6756$ (m),
           \ion{C}{ii} 1334 at $z=1.6359$ (m) \\
PKS1448--232  & \ion{C}{ii} 1334 at $z=1.5855$ (l) \\
Q0122--380          & \ion{Si}{iii} at $z=1.9733$ (m),
        \ion{Si}{ii} 1260 at
        $z=1.9695$ (l), \ion{Si}{iii} at $z=1.9123$ (l),
        \ion{Si}{iii} at $z=1.9102$ (m) \\
Q1101--264          &  \\
\noalign{\smallskip}
\hline
\end{tabular}
\begin{list}{}{}
\item[$^{\mathrm{a}}$]
The letters ``m'' and ``l'' indicate a ``moderate'' uncertainty and a ``low''
uncertainty, respectively. For the class  ``l'' the \ion{H}{i} 
profiles do not change significantly with and without
the uncertain metal lines ({\it cf.} \ion{Si}{ii} 1260 in
Fig.~\ref{fig3} f). The example of the \ion{Si}{iii} absorption 
in Fig.~\ref{fig3} f) is a typical uncertain line fit of  class ``m''.
\end{list}
\end{table*}

\begin{table*}
\caption{The mean PDF of the full sample divided into three and two redshift bins.}
\label{tab_c}
\begin{tabular}{lccccc}
\hline
\noalign{\smallskip}
$F$  & $<z> \, =2.07$  &  $<z> \, =2.52$  & $<z> \, =2.94$  &  $<z> \, =2.41^{a}$ &
 $<z> \, =3.00^{a}$\\
\noalign{\smallskip}
\hline
\noalign{\smallskip}
 0.00 & 0.5700$\pm$ 0.0581 & 0.9189$\pm$ 0.0995 & 1.6960$\pm$ 0.2043  & 0.8009$\pm$ 0.1114 & 1.6277$\pm$ 0.2356 \\
 0.05 & 0.1978$\pm$ 0.0165 & 0.3540$\pm$ 0.0320 & 0.4666$\pm$ 0.0520  & 0.2870$\pm$ 0.0362 & 0.4861$\pm$ 0.0603 \\
 0.10 & 0.1354$\pm$ 0.0115 & 0.2426$\pm$ 0.0190 & 0.2957$\pm$ 0.0312  & 0.1851$\pm$ 0.0227 & 0.3210$\pm$ 0.0396 \\
 0.15 & 0.1401$\pm$ 0.0098 & 0.2064$\pm$ 0.0199 & 0.2799$\pm$ 0.0281  & 0.1664$\pm$ 0.0215 & 0.2950$\pm$ 0.0371 \\
 0.20 & 0.1230$\pm$ 0.0083 & 0.1873$\pm$ 0.0151 & 0.2356$\pm$ 0.0230  & 0.1624$\pm$ 0.0172 & 0.2574$\pm$ 0.0330 \\
 0.25 & 0.1307$\pm$ 0.0078 & 0.2230$\pm$ 0.0206 & 0.2403$\pm$ 0.0212  & 0.1822$\pm$ 0.0177 & 0.2412$\pm$ 0.0291 \\
 0.30 & 0.1216$\pm$ 0.0086 & 0.2003$\pm$ 0.0168 & 0.2385$\pm$ 0.0197  & 0.1630$\pm$ 0.0141 & 0.2350$\pm$ 0.0230 \\
 0.35 & 0.1279$\pm$ 0.0083 & 0.1995$\pm$ 0.0151 & 0.2310$\pm$ 0.0198  & 0.1517$\pm$ 0.0135 & 0.2323$\pm$ 0.0273 \\
 0.40 & 0.1450$\pm$ 0.0088 & 0.2151$\pm$ 0.0133 & 0.2502$\pm$ 0.0195  & 0.1941$\pm$ 0.0146 & 0.2646$\pm$ 0.0283 \\
 0.45 & 0.1488$\pm$ 0.0083 & 0.2574$\pm$ 0.0196 & 0.2496$\pm$ 0.0166  & 0.1907$\pm$ 0.0147 & 0.2448$\pm$ 0.0188 \\
 0.50 & 0.1726$\pm$ 0.0099 & 0.2561$\pm$ 0.0180 & 0.2817$\pm$ 0.0195  & 0.2162$\pm$ 0.0198 & 0.2717$\pm$ 0.0265 \\
 0.55 & 0.1884$\pm$ 0.0104 & 0.2800$\pm$ 0.0170 & 0.3523$\pm$ 0.0201  & 0.2462$\pm$ 0.0207 & 0.3480$\pm$ 0.0235 \\
 0.60 & 0.2209$\pm$ 0.0114 & 0.3323$\pm$ 0.0206 & 0.4246$\pm$ 0.0228  & 0.3158$\pm$ 0.0285 & 0.4009$\pm$ 0.0247 \\
 0.65 & 0.2815$\pm$ 0.0144 & 0.3588$\pm$ 0.0195 & 0.4555$\pm$ 0.0351  & 0.3702$\pm$ 0.0306 & 0.4269$\pm$ 0.0297 \\
 0.70 & 0.3067$\pm$ 0.0136 & 0.4107$\pm$ 0.0183 & 0.5785$\pm$ 0.0342  & 0.4154$\pm$ 0.0254 & 0.5793$\pm$ 0.0432 \\
 0.75 & 0.4045$\pm$ 0.0170 & 0.5631$\pm$ 0.0284 & 0.6888$\pm$ 0.0485  & 0.5417$\pm$ 0.0362 & 0.7354$\pm$ 0.0582 \\
 0.80 & 0.5747$\pm$ 0.0239 & 0.8030$\pm$ 0.0358 & 0.9168$\pm$ 0.0587  & 0.7511$\pm$ 0.0513 & 0.9470$\pm$ 0.0713 \\
 0.85 & 0.8440$\pm$ 0.0317 & 1.1532$\pm$ 0.0712 & 1.4003$\pm$ 0.0660  & 1.0816$\pm$ 0.0731 & 1.4178$\pm$ 0.0733 \\
 0.90 & 1.6501$\pm$ 0.0464 & 2.0285$\pm$ 0.1072 & 2.1911$\pm$ 0.0841  & 2.0036$\pm$ 0.1064 & 2.1899$\pm$ 0.0929 \\
 0.95 & 4.9083$\pm$ 0.0982 & 4.6174$\pm$ 0.1175 & 3.9641$\pm$ 0.1729  & 4.9196$\pm$ 0.1604 & 3.9674$\pm$ 0.1805 \\
 1.00 & 8.6081$\pm$ 0.1986 & 6.1926$\pm$ 0.3069 & 4.5630$\pm$ 0.3659  & 6.6550$\pm$ 0.3420 & 4.5108$\pm$ 0.4290 \\
\noalign{\smallskip}
\hline
\end{tabular}
\begin{list}{}{}
\item[$^{\mathrm{a}}$]
The wavelength ranges for $<z> \, =2.41$ and $<z> \, =3.00$ are 3902--4325 \AA\/ (5 QSOs: HE2347$-$4342, Q0002$-$422,
PKS0329$-$255, Q0453$-$423 and HE1347$-$2457), and 4703--5112 \AA\/ (4 QSOs: Q0055$-$269, PKS2126$-$158,
Q0420$-$388 and and HE0940$-$1050),
respectively.
\end{list}
\end{table*}

\begin{table*}
\caption{The \ion{H}{i} effective optical depth$^{\mathrm{a}}$}
\label{tab_d}
\begin{tabular}{lccccccr}
\hline
\noalign{\smallskip}
QSO & wavelengths (\AA\/) & $<\!z\!>$ &
$\tau_{\ion{H}{i}}^{\mathrm{eff}}$ & $\overline{F}_{\ion{H}{i}}$ & 
$\sigma^{2}_{F}$ & $\tau_{\ion{H}{i}+\mathrm{metal}}^{\mathrm{eff}}$ & Metals (\%)\\
\noalign{\smallskip}
\hline
\noalign{\smallskip}
Q0055--269          & 4785--5577 & 3.262  &  0.412$\pm$0.029 & 0.662$\pm$0.020 & 0.1233$\pm$0.0065 &  0.419$\pm$0.030 & 1.6 \\
                    & 4785--5112 & 3.071  &  0.386$\pm$0.046 & 0.680$\pm$0.033 & 0.1296$\pm$0.0119 &  0.388$\pm$0.046 & 0.5\\
PKS2126--158        & 4638--5112 & 3.010  &  0.263$\pm$0.025 & 0.769$\pm$0.019 & 0.0949$\pm$0.0086 &  0.351$\pm$0.031 & 25.0\\
Q0420--388          & 4231--4909 & 2.759  &  0.308$\pm$0.024 & 0.735$\pm$0.017 & 0.1092$\pm$0.0068 &  0.332$\pm$0.025 & 7.2\\
HE0940--1050        & 4197--4870 & 2.729  &  0.283$\pm$0.023 & 0.754$\pm$0.018 & 0.1004$\pm$0.0086 &  0.325$\pm$0.025 & 13.0\\
HE2347--4342        & 4055--4643 & 2.577  &  0.236$\pm$0.022 & 0.790$\pm$0.018 & 0.0900$\pm$0.0092 &  0.241$\pm$0.022 & 2.3\\
Q0002--422          & 3901--4504 & 2.457  &  0.250$\pm$0.021 & 0.778$\pm$0.016 & 0.0912$\pm$0.0081 &  0.325$\pm$0.025 & 23.1\\
PKS0329--255        & 3815--4439 & 2.395  &  0.183$\pm$0.020 & 0.833$\pm$0.015 & 0.0762$\pm$0.0088 &  0.196$\pm$0.020 & 6.4\\
Q0453--423          & 4084--4362 & 2.474  &  0.253$\pm$0.031 & 0.777$\pm$0.024 & 0.0904$\pm$0.0113 &  0.271$\pm$0.033 & 6.7\\
                    & 3758--3911 & 2.154  &  0.147$\pm$0.026 & 0.864$\pm$0.022 & 0.0523$\pm$0.0110 &  0.161$\pm$0.025 & 9.0\\
HE1347--2457        & 3705--4319 & 2.300  &  0.140$\pm$0.015 & 0.870$\pm$0.013 & 0.0544$\pm$0.0079 &  0.165$\pm$0.016 & 15.4\\
Q0329--385          & 3528--4105 & 2.139  &  0.146$\pm$0.013 & 0.864$\pm$0.012 & 0.0543$\pm$0.0066 &  0.150$\pm$0.013 & 2.7\\
HE2217--2818        & 3509--4091 & 2.126  &  0.138$\pm$0.014 & 0.871$\pm$0.014 & 0.0557$\pm$0.0076 &  0.168$\pm$0.016 & 17.8\\
Q0109--3518         & 3532--4070 & 2.127  &  0.143$\pm$0.015 & 0.867$\pm$0.014 & 0.0590$\pm$0.0078 &  0.168$\pm$0.016 & 14.7\\
HE1122--1648        & 3514--4082 & 2.124  &  0.148$\pm$0.017 & 0.863$\pm$0.014 & 0.0603$\pm$0.0079 &  0.161$\pm$0.017 & 8.0\\
J2233--606          & 3350--3886 & 1.976  &  0.164$\pm$0.019 & 0.849$\pm$0.017 & 0.0707$\pm$0.0106 &  0.179$\pm$0.019 & 8.4\\
PKS0237--23         & 3361--3865 & 1.972  &  0.127$\pm$0.015 & 0.880$\pm$0.013 & 0.0527$\pm$0.0077 &  0.176$\pm$0.019 & 27.6\\
PKS1448--232        & 3306--3860 & 1.947  &  0.145$\pm$0.014 & 0.865$\pm$0.013 & 0.0604$\pm$0.0071 &  0.157$\pm$0.014 & 7.6\\
Q0122--380          & 3282--3819 & 1.921  &  0.133$\pm$0.018 & 0.875$\pm$0.015 & 0.0613$\pm$0.0096 &  0.160$\pm$0.019 & 16.4\\
Q1101--264          & 3503--3765 & 1.989  &  0.111$\pm$0.021 & 0.895$\pm$0.018 & 0.0435$\pm$0.0098 &  0.129$\pm$0.022 & 14.1\\
                    & 3233--3398 & 1.727  &  0.096$\pm$0.020 & 0.908$\pm$0.018 & 0.0316$\pm$0.0110 &  0.131$\pm$0.026 & 26.9\\
\noalign{\smallskip}
\hline
\end{tabular}
\begin{list}{}{}
\item[$^{\mathrm{a}}$]
The errors were estimated using  500 bootstrap realisations of  chunks of 100 pixels (5 \AA\/).
\end{list}
\end{table*}

\begin{table*}
\caption{The evolution of the effective optical depth of the
 full sample divided into redshift bins with width $\Delta z$=0.2$^{\mathrm{a}}$}
\label{tab_e}
\begin{tabular}{llcccllcc}
\hline
\noalign{\smallskip}
$<\!z>$ & &  Wavelengths (\AA\/) &
$\tau_\ion{H}{i}^{\mathrm{eff}}$ & & $<\!z\!>\,$ &  &  Wavelengths (\AA\/) & $\tau_\ion{H}{i}^{\mathrm{eff}}$ \\
\hline
1.80  &                & 3282--3525 &   0.127$\pm$0.013 & & 2.41  &                & 4012--4255 &   0.204$\pm$0.015  \\
      & J2233$-$606    & 3335--3525 &   0.203$\pm$0.039 & &       & HE2347$-$4342  & 4055--4255 &   0.156$\pm$0.022 \\
      & PKS0237$-$23   & 3361--3525 &   0.111$\pm$0.024 & &       & Q0002$-$422    & 4012--4255 &   0.228$\pm$0.034 \\
      & PKS1448$-$232  & 3306--3525 &   0.126$\pm$0.022 & &       & PKS0329$-$255  & 4012--4255 &   0.186$\pm$0.034 \\
      & Q0122$-$380    & 3282--3525 &   0.097$\pm$0.018 & &       & Q0453$-$423    & 4084--4255 &   0.252$\pm$0.042 \\
      & Q1101$-$264    & 3282--3398 &   0.095$\pm$0.027 & &       & HE1347$-$2457  & 4012--4255 &   0.205$\pm$0.022  \\
2.00  &                & 3525--3769 &   0.141$\pm$0.008 & & 2.59  &                & 4255--4498 &   0.255$\pm$0.016 \\
      & Q0329$-$385    & 3528--3769 &   0.128$\pm$0.017 & &       & Q0420$-$388    & 4255--4498 &   0.266$\pm$0.031 \\
      & HE2217$-$2818  & 3525--3769 &   0.133$\pm$0.023 & &       & HE0940$-$1050  & 4255--4498 &   0.293$\pm$0.033 \\
      & Q0109$-$3518   & 3532--3769 &   0.156$\pm$0.026 & &       & HE2347$-$4342  & 4255--4498 &   0.258$\pm$0.022 \\
      & HE1122$-$1648  & 3525--3769 &   0.117$\pm$0.019 & &       & Q0002$-$422    & 4255--4498 &   0.268$\pm$0.034 \\
      & J2233$-$606    & 3525--3769 &   0.168$\pm$0.039 & &       & PKS0329$-$255  & 4255--4439 &   0.170$\pm$0.037 \\
      & PKS0237$-$23   & 3525--3769 &   0.146$\pm$0.022 & &       & Q0453$-$423    & 4255--4362 &   0.255$\pm$0.060 \\
      & PKS1448$-$232  & 3525--3769 &   0.157$\pm$0.021 & & 2.80  &                & 4498--4741 &   0.297$\pm$0.024 \\
      & Q0122$-$380    & 3525--3769 &   0.165$\pm$0.017 & &       & PKS2126$-$158  & 4638--4741 &   0.231$\pm$0.052 \\
      & Q1101$-$264    & 3525--3765 &   0.098$\pm$0.021 & &       & Q0420$-$388    & 4498--4741 &   0.308$\pm$0.034 \\
2.20  &                & 3769--4012 &   0.147$\pm$0.009 & &       & HE0940$-$1050  & 4498--4741 &   0.304$\pm$0.033 \\
      & Q0002$-$422    & 3901--4012 &   0.269$\pm$0.046 & &       & HE2347$-$4342  & 4498--4643 &   0.315$\pm$0.025 \\
      & PKS0329$-$255  & 3815--4012 &   0.191$\pm$0.036 & & 2.99  &                & 4741--4984 &   0.338$\pm$0.024 \\
      & Q0453$-$423    & 3769--3911 &   0.142$\pm$0.022 & &       & Q0055$-$269    & 4785--4984 &   0.389$\pm$0.062 \\
      & HE1347$-$2457  & 3769--4012 &   0.094$\pm$0.021 & &       & PKS2126$-$158  & 4741--4984 &   0.310$\pm$0.044 \\
      & Q0329$-$385    & 3769--4012 &   0.138$\pm$0.017 & &       & Q0420$-$388    & 4741--4914 &   0.380$\pm$0.038 \\
      & HE2217$-$2818  & 3769--4012 &   0.152$\pm$0.022 & &       & HE0940$-$1050  & 4741--4870 &   0.260$\pm$0.046 \\
      & Q0109$-$3518   & 3769--4012 &   0.127$\pm$0.027 & & 3.18  &                & 4984--5227 &   0.304$\pm$0.035 \\
      & HE1122$-$1648  & 3769--4012 &   0.163$\pm$0.019 & &       & Q0055$-$269    & 4984--5227 &   0.362$\pm$0.050 \\
      & J2233$-$606    & 3769--3886 &   0.096$\pm$0.019 & &       & PKS2126$-$158  & 4984--5112 &   0.203$\pm$0.059  \\
      &                &            &                   & & 3.44$^{b}$ & Q0055$-$269  & 5227--5577 &   0.462$\pm$0.045\\
\noalign{\smallskip}
\hline
\end{tabular}
\begin{list}{}{}
\item[$^{\mathrm{a}}$]
The errors were estimated using  500 bootstrap realisations of chunks of 100 pixels (5 \AA\/).
For each bin, \eff\/ was estimated from the mean flux of all
pixels from the QSOs listed.
\item[$^{\mathrm{b}}$]
Q0055$-$269 is the only QSO in this $z$ bin. 
\end{list}
\end{table*}

\end{document}